\documentclass[journal]{IEEEtran}
\ifCLASSINFOpdf
  \usepackage[pdftex]{graphicx}
  \graphicspath{../eps/}
  \DeclareGraphicsExtensions{.pdf,.jpeg,.png}
\else
  \usepackage[dvips]{graphicx}
  \graphicspath{{../eps/}}
  \DeclareGraphicsExtensions{.eps}
\fi
\usepackage{siunitx}
\usepackage[colorlinks=false]{hyperref}
\usepackage[xcdraw]{xcolor}
\usepackage{cite}
\usepackage{amsmath}
\usepackage{amssymb}
\usepackage{mathptmx}
\usepackage{epsfig}
\usepackage{times}
\usepackage{threeparttable}
\usepackage{stfloats}
\usepackage{amsmath,amssymb,amsfonts}
\usepackage{algorithmic}
\usepackage{textcomp}
\usepackage{array}
\usepackage{multirow}
\usepackage{makecell}
\usepackage{comment}
\usepackage{array}

  \usepackage[caption=false,font=footnotesize]{subfig}
\begin{document}
%
\title{Design-Agnostic Distributed Timing Fault Injection Monitor with End-to-End Design Automation}
%
%
%

\author{
        Yan He*,~\IEEEmembership{Graduate Student Member,~IEEE,}
        Yumin Su*,~\IEEEmembership{Graduate Student Member,~IEEE,}
        and~Kaiyuan~Yang,~\IEEEmembership{Member,~IEEE}
        \thanks{Manuscript received on}
\thanks{(Yan He and Yumin Su contributed equally to this paper.)}
\thanks{Y. He, Y. Su, and K. Yang are with the Department of Electrical and Computer Engineering, Rice University, Houston TX, 77005, USA. (Corresponding Author: Kaiyuan Yang, kyang@rice.edu)}
\thanks{1063-8210 © 2023 IEEE. Personal use is permitted, but republication/redistribution
requires IEEE permission.}
}

%
%

\markboth{IEEE Journal of Solid-State Circuits}%
{Shell \MakeLowercase{\textit{et al.}}: Bare Demo of IEEEtran.cls for IEEE Journals}
%



\maketitle

\begin{abstract}

Fault Injection Attacks (FIAs) induce hardware failures in circuits and exploit these faults to compromise the security of the system. It has been demonstrated that FIAs can bypass system security mechanisms, cause faulty outputs, and gain access to secret information. Certain types of FIAs can be mounted with little effort by tampering with clock signals and/or the chip's operating conditions. To mitigate such low-cost yet powerful attacks, we propose a fully synthesizable and distributable \textit{in situ} Fault Injection Monitor that employs a Delay Locked Loop (DLL) to track the pulse width of the clock. We further develop a fully automated design framework to optimize and implement the FIA monitors at any process node. 
Our design is fabricated and verified in 65nm CMOS technology with a small footprint of 1500$\boldsymbol{\mu}\textbf{m}^\textbf{2}$. It can lock to clock frequencies from 2MHz to 1.26GHz while detecting all twelve types of possible clock glitches, as well as timing FIA injections via the supply voltage, electromagnetic signals, and chip temperature. 

\end{abstract}


%
\IEEEpeerreviewmaketitle

\section{Introduction}

Fault Injection Attacks (FIAs) induce errors in the logic operations of a chip and exploit such errors to break systems that are otherwise secure. As an example, when a system is validating some information, a hardware failure could trick the system into accepting incorrect data. All subsequent operations that depend on the authenticity of this information would become insecure because of this injected fault. 
The effectiveness and risk of FIAs have been established through academic research and real-world attacks. Multiple gaming consoles have been hacked by introducing turbulence to the clock or the voltage supply \cite{giller_implementing_2015, lu_injecting_2019}. Drones were attacked without any physical contact: electromagnetic (EM) interference can inject faults into drones remotely \cite{gonzalez_drone_2023}. Even CPU functions carefully designed for security have shown vulnerabilities in the face of FIAs. \cite{murdock_plundervolt_2020} breaks Intel SGX by injecting faults with software, while \cite{qiu_voltjockey_2019} attacks ARM TrustZone by manipulating the processor's voltage. The fact that FIAs open up backdoors to flawlessly designed secure systems makes them a particularly devastating threat to modern computing systems. 

As evident from the examples above, FIAs can be launched with a variety of methods and tools. But, broadly speaking, FIAs exploit two major categories of physical anomalies: directly flipping the voltage of a signal wire or a memory/register unit, and violating timing constraints to make a register sample a wrong value. 

\textbf{Bit Flips.} Laser is the most powerful tool to induce bit flips and cause highly localized faults in the chip. It is even possible to attack individual gates and wires with high-power lasers. While laser attacks are powerful, mounting such attacks requires expensive equipment and considerable complexity. More importantly, it has been demonstrated in various works that laser attacks can be detected with low-cost light sensors densely distributed across the chip. \cite{matsuda_286_2018} reports a 28\% increase in area when the AES module is protected with Bulk Built-In Current Sensors, while the photosensors in \cite{zhang_laser_2023} achieve an area overhead down to 20\%. \cite{kumar_100gbps_2023} employs standard inverters to form a laser detection circuit array, whose area overhead is reported to be as low as 3\% at an advanced node. Considering the gap between the requirements for attackers and defenders, we argue that these protections are sufficient to thwart most laser-based attacks. 

\textbf{Timing Violations.} Logic faults can also be injected through intentional timing violations. The attacker alters the relative timing relationship between the clock and the data arrival time, causing the register to latch a wrong value. Timing FIAs can be conducted by directly hijacking the clock signal or by indirectly influencing the logic gates' delay and, thus, data arrival time. As the delay of a CMOS gate can be affected by various environmental factors (voltage, electromagnetic interference, temperature, etc.), a broad attack surface exists for timing FIAs. More importantly, simple tools and even software hacks can be exploited to interfere with these factors and launch timing FIAs. Since timing FIAs are inexpensive yet effective, we focus on mitigating this type of FIA in this work. We aim to develop low-cost timing FIA sensors that will complement the aforementioned laser sensors to cover a broad spectrum of FIAs. 


An ideal solution to FIAs should be applicable to any digital design, be compatible with the typical digital design workflow in any technology node, and require minimum design and testing efforts. 
Specifically, we target a \textit{fully synthesizable} FIA sensor design that can be generated by a compiler for synthesis and placement \& routing (P\&R), just like standard digital circuits. It should also handle the parasitics and layout mismatches introduced in the automatic P\&R process without post-silicon testing. 
Furthermore, the hardware development costs will be greatly reduced if the FIA mitigation can be \textit{automatically generated} by a framework. 
The user only needs to provide the process design kit (PDK), the standard cell library, and a few parameters regarding the system's clock. The framework can then run the required simulations and provide a netlist ready for layout with electronic design automation (EDA) tools. As such, the framework will enable agile and low-cost insertion of FIA mitigation to any existing design at any technology node. 

To achieve these goals, we present a fully synthesizable design-agnostic timing FIA monitor along with an automatic design framework. The principle of the monitor is to replicate the system clock with an internal delay-locked loop (DLL) and compare it with every clock edge in real time. Our monitor is able to detect attacks on the clock signal as well as the logic gates' propagation/contamination delay. It features a tiny footprint using standard cells and can be distributed across the chip for localized FIA protection. Our prototype of the timing FIA monitor in 65nm CMOS demonstrates:
\begin{itemize}
    \item robust detection of timing faults, supporting a wide range of clock frequencies from $2MHz$  to $1.26GHz$; 
    \item a fully synthesized PVT-robust design validated across the automotive temperature range (-40 to $125^\circ C$), 0.5 to $1.4V$ VDD, and 50 DUTs;
    \item an end-to-end automation framework to implement the monitor in any technology node;
    \item low power consumption (0.2 - $1.12mW$) for 2 - $1250MHz$ clocks at 1.2V, $25^\circ C$ and small footprints ($1500\mu m^2$).
\end{itemize}

This article extends \cite{he_synthesizable_2024} and is organized as follows. Section~\ref{timing_fia} summarizes the attack surface and possible mitigations of timing FIAs. Section~\ref{FIA_monitor} elaborates on our proposed monitor, followed by a description of the automation framework in Section~\ref{gen_framework}. Section~\ref{measure_results} offers the measurement results of 65nm test chips, while Section~\ref{tech_scale} showcases the design framework in 28nm. Finally, Section~\ref{conclusion} concludes the article.

\section{Timing Fault Injection Attacks and Mitigations}
\label{timing_fia}

\begin{figure}[t]
  \centering
  \includegraphics[width=\linewidth]{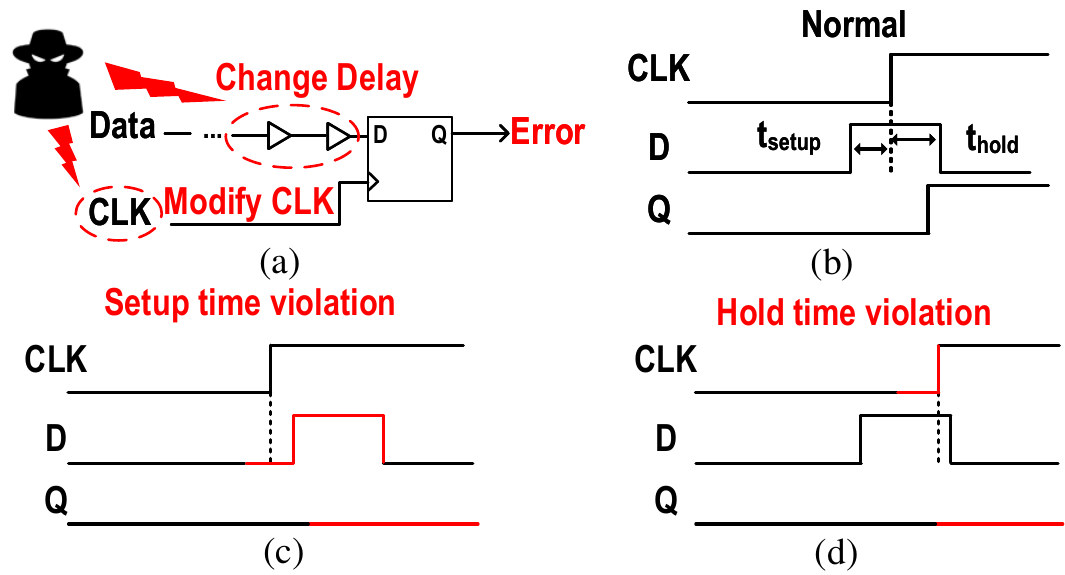}
  \vskip -2ex
  \caption{(a) Timing FIAs entry points: the clock signal and the gates' delay. (b) Correct timing at the register. (c) The case where the data arrives too late, and (d) the case where the data changes too fast. }
  \label{reg_timing}
\end{figure}

For standard synchronous digital circuits to function correctly, it is essential that all timing constraints are satisfied. The timing at the registers is particularly critical because an incorrectly sampled value cannot be recovered or even detected, leading to system failures and logic errors that attackers can exploit. The \textit{setup time} requirement of a register refers to how much earlier the data signal must be ready \textit{before} the clock edge, while the \textit{hold time} requirement states that the data signal cannot change for a certain amount of time \textit{after} the clock edge. Violating either of them will lead to a timing failure and potentially wrong values sampled by the register. 
Timing FIAs induce faults by intentionally breaching the timing requirements, as shown in Fig.~\ref{reg_timing}, by targeting either the clock signal or the data arrival time. 

\subsection{Clock Glitching}
Hijacking the clock signal is the most intuitive way to trigger a timing violation since the setup time and the hold time requirements are relative to the clock edge. In a highly optimized design, shifting the clock signal forward/backward will likely violate the setup/hold time requirement \cite{yang_design_2020}, \cite{lashermes_dfa_2012}, \cite{ning_modeling_2018}, \cite{matsubayashi_clock_2016}. Clock glitches and abnormal clock patterns can be induced through external clock pins, software/firmware control of the clock generator, physical interference, or hardware trojans. More advanced attacks have been demonstrated recently, in which dynamic voltage and frequency scaling of a CPU/GPU is exploited to disrupt the timing~\cite{sun_lightning_2023}. 

\begin{figure}[t]
  \centering
  \includegraphics[width=0.95\linewidth]{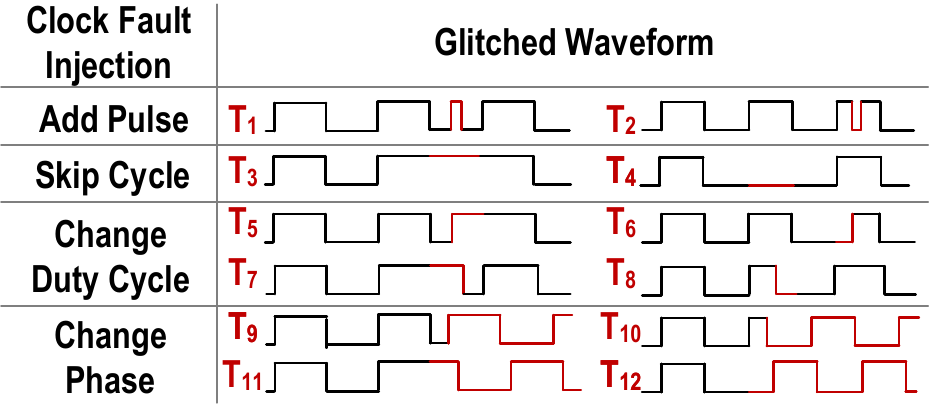}
  \vskip -2ex
  \caption{A summary of twelve possible types of clock glitches. }
  \label{glitch_type}
\end{figure}

Apparently, a wide variety of clock glitches can be induced in practice. Without losing generality, we categorize all possible clock glitches into four major classes with twelve types of waveforms (Fig.~\ref{glitch_type}) to illustrate the coverage of our FIA sensor. While these elemental glitches may or may not form a valid FIA when used alone, real-world clock glitches are likely a temporal-spatial combination of the elemental glitches and will be detected if our sensor covers all the basic elements. 
\subsubsection{Pulse Addition} 
Additional pulses can be added to the clock signal. $\text{T}_1$ represents the scenario where an additional pulse is inserted into the negative phase of the clock, while $\text{T}_2$ shows an added pulse at the positive phase. 
\subsubsection{Cycle Skipping}
Clock pulses can also be removed. A positive (negative) clock phase is skipped in $\text{T}_3$ ($\text{T}_4$). 
\subsubsection{Duty Cycle Change}
Attackers may choose to modify the duty cycle in one clock period as well. $\text{T}_5$ and $\text{T}_7$ extend the next and the previous positive phase, respectively. Similarly, $\text{T}_6$ and $\text{T}_8$ increase the negative phase. 
\subsubsection{Phase Shift}
The clock phase may be shifted by the attacker starting from an arbitrary time point. The attacker shifts the phase forward when the clock is at the low (high) voltage in $\text{T}_9$ ($\text{T}_{10}$). Likewise, the clock phase is shifted backward in $\text{T}_{11}$ and $\text{T}_{12}$. Note that this class of clock glitches changes only one positive or negative phase. This characteristic differs from the change duty cycle class, which varies both the positive and the negative phases. 


\subsection{Delay Manipulation of the Logic Paths}
\begin{figure}[t]
  \centering
  \includegraphics[width=\linewidth]{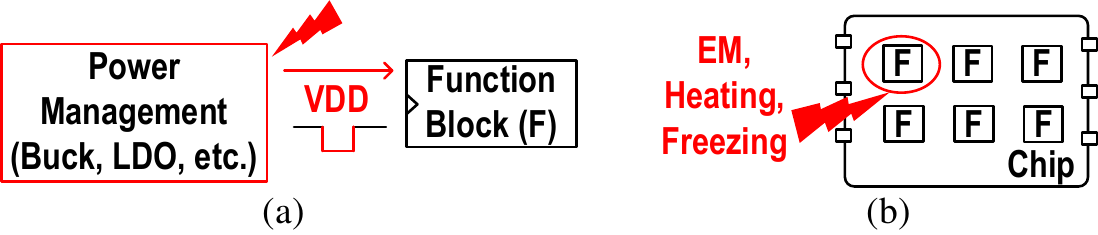}
  \vskip -2ex
  \caption{To alter the logic gates' delay, an attacker may (a) fault the power management circuit to produce a supply voltage glitch, or (b) remotely inject EM signals or temperature changes to a localized region of the chip. }
  \label{delay_manipulate}
\end{figure}

The other entry point for timing FIAs is the data input to the register. If the clock edge remains the same but the data arrives too late, a setup time violation will happen. Similarly, a hold time violation can be triggered when the data changes too fast after the clock edge. 
An attacker can hack the delay of logic gates and thus the data arrival time, by altering the chip's operating conditions, as shown in Fig.~\ref{delay_manipulate}. 

\subsubsection{Voltage}
The delay of the logic gates is highly dependent on the supply voltage. As the supply voltage decreases, the delay will increase. \cite{lu_injecting_2019} applies this technique to bypass a read length assertion in PlayStation Vita to exploit a buffer overflow vulnerability. ChipWhisperer \cite{noauthor_chipwhisperer_nodate}, an open-source toolchain, is used to perform the voltage glitching attack in this work. This readily available and affordable toolchain underscores the versatility of timing FIAs. 

\subsubsection{Electromagnetic Coupling}
Other than controlling the supply voltage directly, the attacker may choose to remotely couple a high-power EM signal to the power rail. Short electromagnetic pulses (EMP) have been demonstrated to trigger both data-dependent and constant faults during Advanced Encryption Standard (AES) computations in \cite{dehbaoui_electromagnetic_2012}. EMP attacks affect the circuit's delay in a way similar to voltage glitch attacks: the supply voltage is suddenly lowered for a short interval and causes the delay to increase temporarily. More subtle attacks can be achieved through electromagnetic interference (EMI)~\cite{hayashi_transient_2013}. EMI enables the attacker to precisely interfere with the security module by coupling noise to the circuit's supply cable at a frequency that has less attenuation for the security module. \cite{fujimoto_detection_2018} injects into the device an EM interference whose amplitude is as small as $90mV$. The attack requires a longer time to take effect, but the voltage disturbance on the victim chip's power supply is smaller. The small voltage difference makes EMI attacks more difficult to detect. 

\subsubsection{Temperature}
It is well known that the delay of CMOS gates is sensitive to temperature variations, which offers another entry point for FIAs. High temperatures also damage the data stored in the memories. For example, \cite{hutter_temperature_2014} reports fault injection in the RSA encryption by heating the hardware. The faulty RSA computation is then exploited to reveal the secret RSA primes, showcasing the damaging consequence of FIAs.

It is worth noting that EM and temperature attacks targeting data paths or clock trees can be highly localized. They need not affect the functionality of the rest of the chip. This necessitates distributed low-cost sensors to protect the full chip. 

\subsection{Existing Timing Fault Mitigations}
Generally, chip-level FIA mitigations follow three directions: logic checking, adaptive design, and anomaly detection. 

\subsubsection{Logic Checking} 
Hardware-enforced assertions can be integrated into the chip to detect faults. \cite{kumar_100gbps_2023} employs a combination of parity and algorithmic checks to ensure that the computations inside the AES block produce the expected results. This method is effective against any AES fault but is not directly applicable to other designs. In order to capture FIAs in other circuits, all the checkers must be redesigned, which takes considerable engineering effort. Pre-silicon logic analysis can also be performed to identify circuits that are susceptible to FIAs \cite{nasahl_synfi_2022, shuvo_ldtfi_2022}. These frameworks try to resolve FIAs at design time and can reduce manufacturing and testing costs. They still, unfortunately, require human experts to input security properties or critical registers that are highly design-specific. In addition, logic checking is limited to specialized accelerators, since the flexibility of the logic executed on general-purpose processors makes it extremely challenging to predict and verify the circuit's outputs with low overheads. 

\subsubsection{Adaptive Design}
Timing adaptive design is a well-studied topic aiming to adapt the voltage and/or clock of a digital system based on the specific process, voltage, temperature (PVT) condition of a chip. Adaptive designs are generally achieved through either PVT sensors, or in-situ error detectors. As such, their principles are highly relevant to timing FIA detection and thus can potentially be reused here. We will discuss two representative techniques and their pros and cons as FIA monitors. 
First, Razor is an adaptive technique based on error detection, which looks for setup time violations \cite{ernst_razor_2003}, \cite{das_razorii_2009}, \cite{kwon_razor-lite_2014}, \cite{zhang_irazor_2018}. Tiny transition detection circuits, as small as three extra transistors~\cite{zhang_irazor_2018}, are embedded in critical registers/latches to detect data changes within a speculation window right after the clock edge. Therefore, Razor is supposed to detect timing FIAs that induce setup time violations. 
However, Razor techniques always face the challenge of deciding whether a data transition after the clock edge is a delayed arrival from the previous cycle or a fast transition in the current cycle. Razor avoids this ambiguity by enforcing a short-path constraint: all paths to a Razor register must be buffered to meet a minimum contamination delay. 
To adapt to PVT variations, the minimum delay can be relatively small, and only a very small portion of registers on critical paths need to be equipped with Razor. If Razor is employed to detect intentional clock glitches, the detection window will become so large that the buffers' overhead will be significant. Additionally, some clock glitches, such as cycle skipping and phase shifts, cannot be detected by Razor. 
The Tunable Replica Circuit (TRC) is a well-known canary-type delay replica based on a programmable delay line calibrated to the delay of the critical path \cite{cho_postsilicon_2017, bowman_22_2013}. 
As timing FIAs often alter the gate delays, the TRC can identify malicious changes in the clock signal or operating conditions by detecting delay changes in the replica circuits. \cite{nemiroff_fault-injection_2022} discusses the feasibility of reusing TRC to detect timing FIAs. However, most delay replica designs in prior works require post-silicon calibration. To serve as timing FIA monitors, they must be distributed across the chip in much higher quantities, leading to high power and area overheads, as well as testing costs. Moreover, a time-to-digital converter with crafted pattern-matching logic is necessary to cover all types of glitches, further increasing the overhead. In summary, adaptive design techniques were developed with highly relevant but different goals as timing FIA monitors. They inspire the design of FIA countermeasures but are not optimal or comprehensive by themselves. 

\subsubsection{Anomaly Detection} 
The third class of FIA protection is by detecting physical anomalies in the chip. For example, clock glitches can be directly monitored using an on-chip oscilloscope that oversamples the system clock. In \cite{song_fll-based_2022}, an FLL is designed to lock to the clock's frequency and monitor the clock waveform in the following cycles. While this method can recognize all types of clock glitches, the FLL's large area and power consumption hinder the monitor from being distributed across the chip to cover localized FIAs. Additionally, this method requires a high oversampling ratio so that the supported system clock frequency is severely restricted. It is also dedicated to the clock waveform with low protection for data arrival time. 
As another example, \cite{miura_local_2014, seo_pg-cas_2021} developed an on-chip sensor to detect the electromagnetic probes that are placed near the chip to launch EM attacks. The sensor employs an on-chip LC oscillator whose frequency shift indicates the existence of abnormal coupling to the on-chip inductor. These sensors excel at detecting one particular type of EM attack but cannot generalize to other types of timing FIAs.


\section{DLL-Based Synthesizable FIA Monitor}
\label{FIA_monitor}
We present a design-agnostic, compact, and distributable FIA monitor based on anomaly detection. 
Despite its small footprint, it detects both clock glitching and delay manipulation attacks. 
The monitor is a fully synthesizable soft IP that can be easily integrated into any existing design. Its negligible design and testing costs, together with minimal power and area overhead, enable the monitor to be distributed across the chip, providing extensive coverage for localized attacks.

\subsection{Timing Anomaly Detection with Clock Replica}
\begin{figure}[t]
  \centering
  \includegraphics[width=\linewidth]{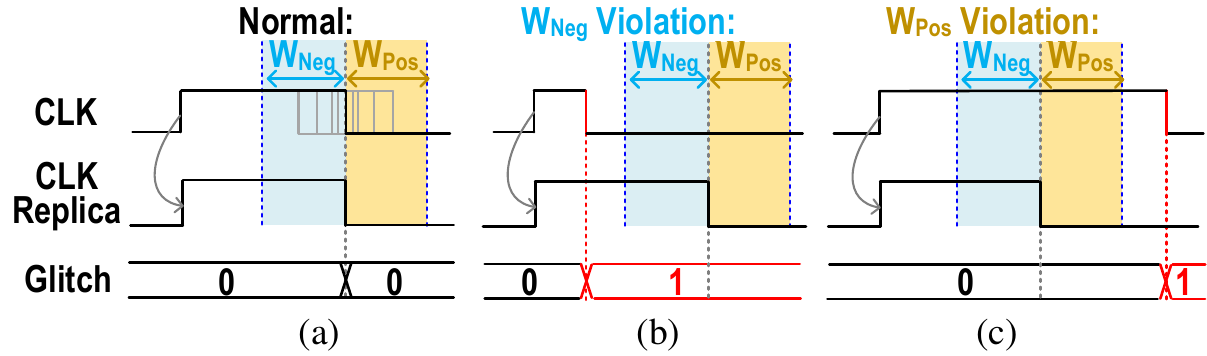}
  \vskip -2ex
  \caption{(a) $\text{W}_\text{Neg}$ and $\text{W}_\text{Pos}$ determine the acceptance window and normal clock jitters should fall inside this window. (b) and (c) depict the scenarios where the clock's pulse width is shorter or longer than the expected value.}
  \label{clock_replica}
\end{figure}

The key principle of our FIA monitor is to track the clock pulse width with a clock replica circuit. The clock replica circuit locks its pulse width to that of the normal system clock. Since any clock signal naturally exhibits jitter and variations from noise and uncertainties, we introduce a programmable acceptance window, as shown in Fig.~\ref{clock_replica}. Under normal operations, the system clock falls within this acceptance window. If the clock's pulse width is shorter (longer) than the programmed minimum (maximum) threshold, the replica circuit reports an attack has been detected.

\begin{figure}[t]
  \centering
  \includegraphics[width=\linewidth]{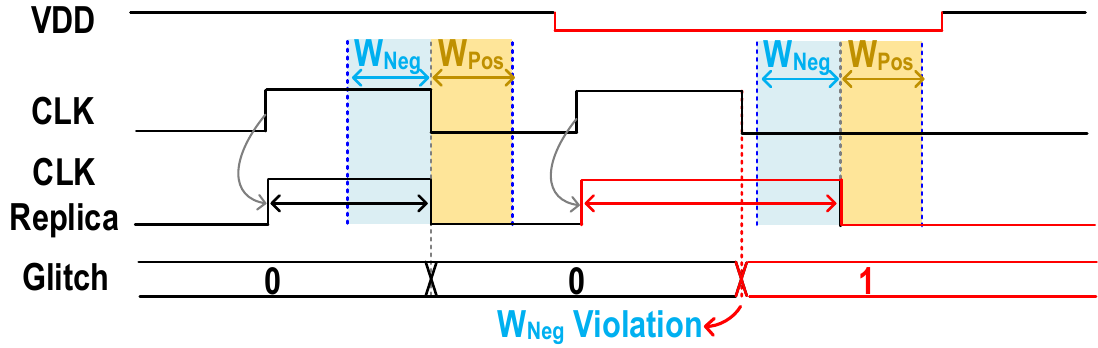}
  \vskip -2ex
  \caption{A voltage glitch attack increases the delay and causes the clock's pulse width to be smaller than the minimum expected value.}
  \label{supply_attack}
\end{figure}

This clock pulse width monitoring scheme is effective regardless of whether the attacker hijacks the clock signal or alters the data path delay. If an FIA is mounted by altering the clock signal, the system clock's pulse width will differ from the locked pulse width and the attack can be detected. If the attacker targets timing violations by glitching the voltage, temperature, and EM conditions, the distributed clock replica's pulse width will be affected by the attack and deviate from the actual clock (Fig.~\ref{supply_attack}), thus raising alerts. The only possible way to bypass the monitor is to change the delay of the gates and then match that delay with the pulse width of the clock accordingly. However, since the delay and the clock match in this case, there are no timing violations in the first place.

\subsection{Same-Cycle FIA Alert Generation}

\begin{figure}[t]
  \centering
  \includegraphics[width=\linewidth]{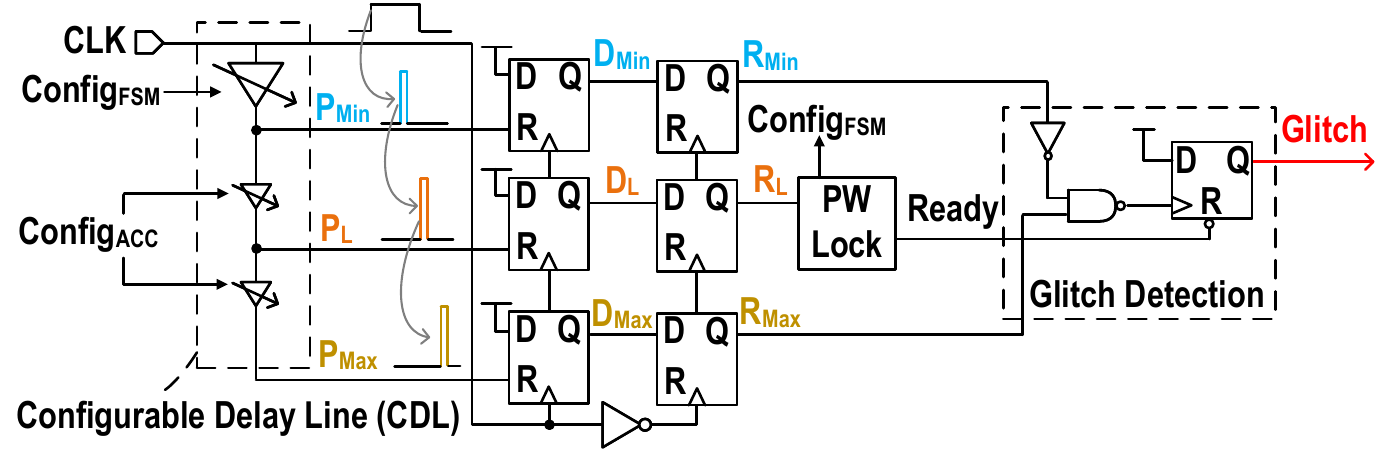}
  \vskip -2ex
  \caption{Implementation of the clock replica and the acceptance window.}
  \label{architecture}
\end{figure}

We employ digital delay lines to realize the clock replica and the acceptance window, as shown in Fig.~\ref{architecture}. The clock is first passed into the main configurable delay line to produce a pulse $\text{P}_\text{Min}$, which determines the minimum acceptable delay $\text{D}_\text{Min}$. A second configurable delay line is used to further delay the rising edge of the clock, creating $\text{P}_\text{L}$. $\text{D}_\text{L}$, derived from $\text{P}_\text{L}$, will be calibrated by an FSM to be the same as the clock's pulse width. Similarly, the maximum acceptable clock uncertainty is $\text{D}_\text{Max}$, which is set by $\text{P}_\text{Max}$, the output of the last configurable delay line. 

\begin{figure}[t]
  \centering
  \includegraphics[width=\linewidth]{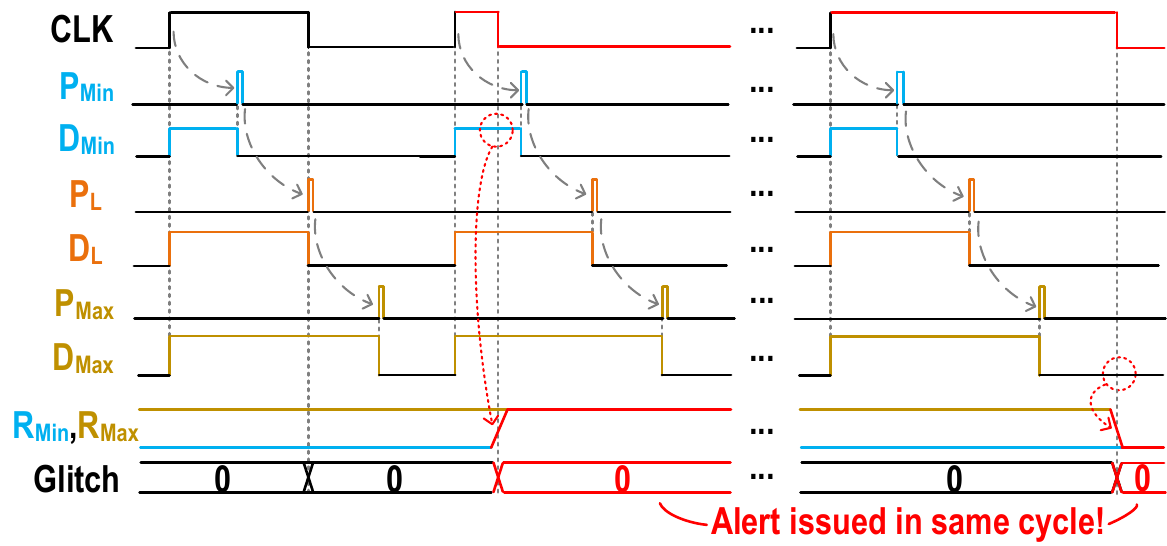}
  \vskip -2ex
  \caption{When the monitor locks to the clock, $\text{P}_\text{L}$ is aligned to the clock's falling edge. An FIA alert is issued in the same cycle if $\text{R}_\text{Min}\neq0$ or $\text{R}_\text{Max}\neq1$.}
  \label{alert_gen}
\end{figure}

With the three pulses generated by the configurable delay lines, FIAs can be detected in the same cycle as the fault is injected (Fig.~\ref{alert_gen}). The values of $\text{D}_\text{Min}$, $\text{D}_\text{L}$, and $\text{D}_\text{Max}$ are sampled at the falling edge of the clock signal to produce $\text{R}_\text{Min}$, $\text{R}_\text{L}$, and $\text{R}_\text{Max}$. Under normal operations, the clock's negative edge comes after $\text{D}_\text{Min}$ but before $\text{D}_\text{Max}$. Therefore, $\text{R}_\text{Min}$ should be 0 while $\text{R}_\text{Max}$ needs to be 1. Any other combination of $\text{R}_\text{Min}$ and $\text{R}_\text{Max}$ values indicates that the clock pulse width diverges from the expected value and an alert is raised.

\begin{figure}[t]
  \centering
  \includegraphics[width=\linewidth]{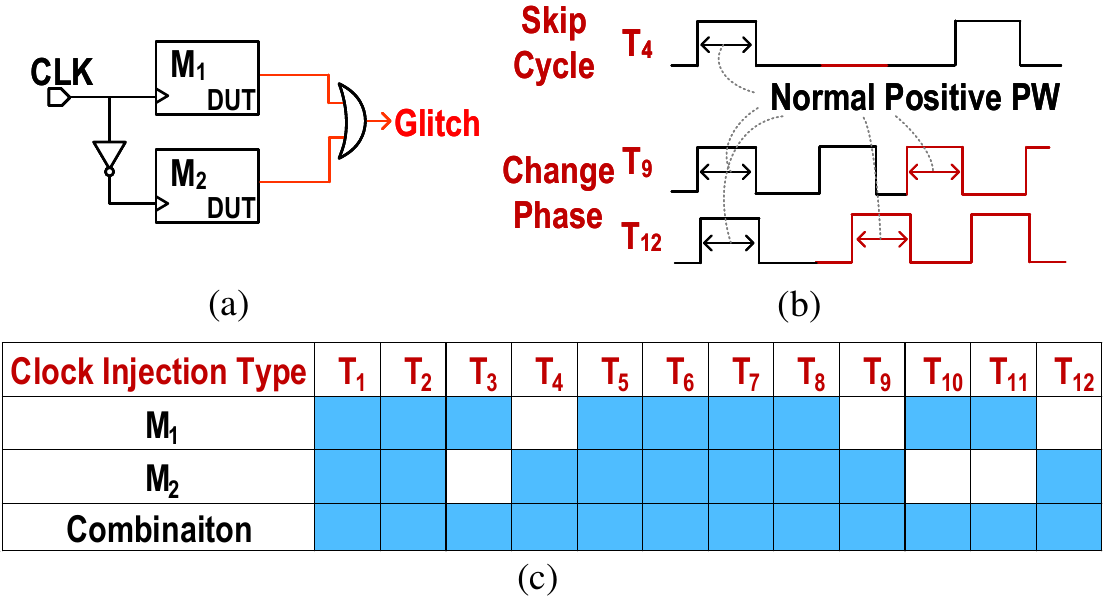}
  \vskip -2ex
  \caption{(a) A pair of monitors are required to detect all clock glitch types. (b) Undetected clock glitches with $\text{M}_1$ alone. (c) Summary of detectable glitches.}
  \label{monitor_pair}
\end{figure}

A single FIA monitor is able to detect nine out of twelve types of clock glitches because the monitor only locks and monitors the positive pulse width of the clock. The remaining three glitch types affect the negative pulse width of the clock while leaving the positive pulse width intact, as illustrated in Fig.~\ref{monitor_pair}b. To capture these glitches, another copy of the monitor can be inserted to track the negative phase of the clock. With this dual-monitor setup, all types of glitches can be captured. The placement of these monitors is flexible; they are not required to be deployed in pairs. Each monitor can work independently to detect most types of clock glitches as well as voltage, EM, and temperature glitches. 
Alternately placing the monitors for the positive and the negative clock phases across the chip in a checkerboard fashion is a possible scheme for increasing comprehensive coverage with low overheads. 

\subsection{Digitally Controlled Delay Line}
To ensure that the FIA monitor is synthesizable, the delay lines are constructed using only standard cells. This standard-cell-only design enables automatic place and route with software and guarantees that the monitor is technology-agnostic. 

\begin{figure}[t]
  \centering
  \includegraphics[width=\linewidth]{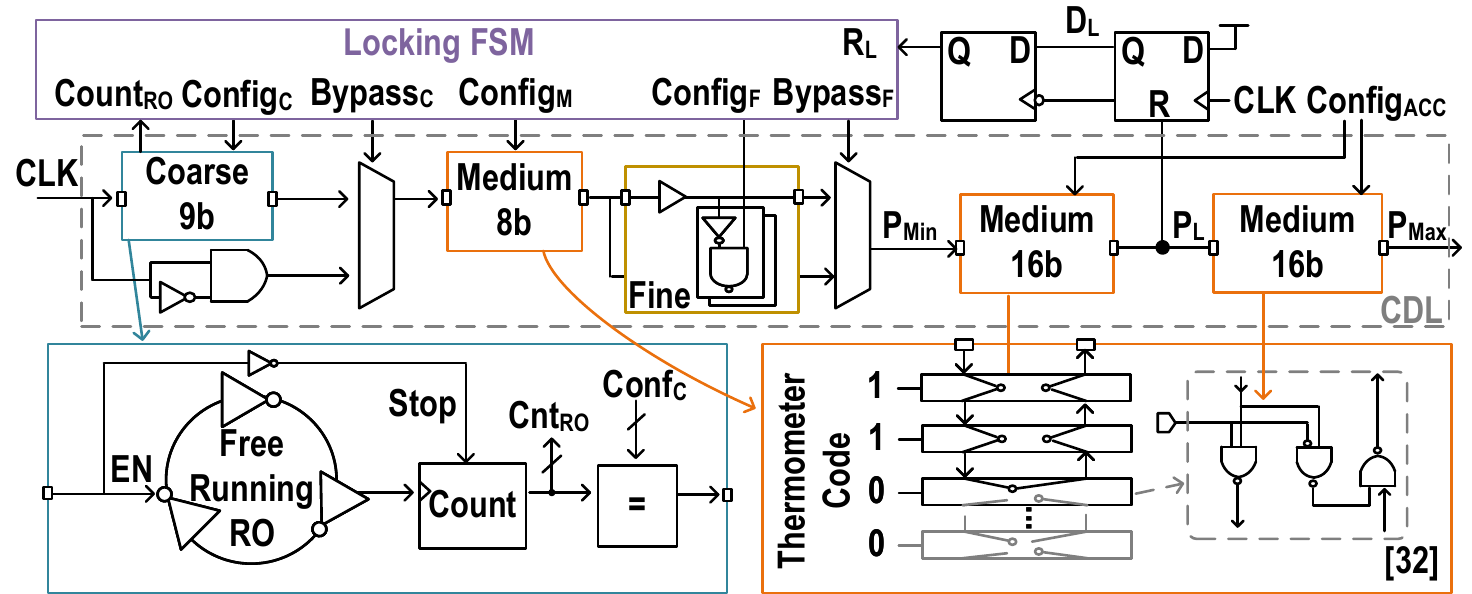}
  \vskip -2ex
  \caption{Schematic of the Configurable Delay Line (CDL).}
  \label{delay_line}
\end{figure}

The main configurable delay line is broken down into three stages in order to provide a wide delay tuning range while maintaining a small footprint (Fig.~\ref{delay_line}). The coarse stage counts the cycles of a ring oscillator (RO). Once the pre-defined cycle number ($\text{Conf}_\text{c}$) has been reached, the coarse stage outputs 1, and the medium stage takes over. The medium stage is a path-selection delay line with thermometer-coded control bits \cite{liu_fully_2021}. When the control bits are all one, the input needs to propagate through all the stages to reach the output port. If the control code is all zero, on the other hand, the input travels through the first stage and goes directly to the output. Note that the two delay lines to tune the acceptable window also have the same structure. The fine stage is built from standard-cell-based varactors with sub-gate-delay resolution. Such varactors work on the principle that the capacitance at a logic gate's port (NAND, NOR, etc.) depends on the voltage at other ports. By holding other ports at 0 or 1, the delay, which is affected by the loading capacitance, can be tuned at a high precision. 

Here, both the coarse stage and the fine stage can be bypassed for two purposes: to increase the maximum locking frequency and to reduce power consumption. The coarse stage and the fine stage have a relatively large delay offset, i.e., the inherent delay when the control code is 0. When all control bits are 0, skipping these two stages can further reduce the delay and increase the upper bound of the frequency range at the cost of resolution loss. As for the power consumption, the coarse stage (RO plus the counter) dominates the total power usage. Bypassing the coarse stage prevents the RO from oscillating and, therefore, reduces the power consumption. 
  

\subsection{Automatic Pulse Width Locking}

\begin{figure}[t]
  \centering
  \includegraphics[width=\linewidth]{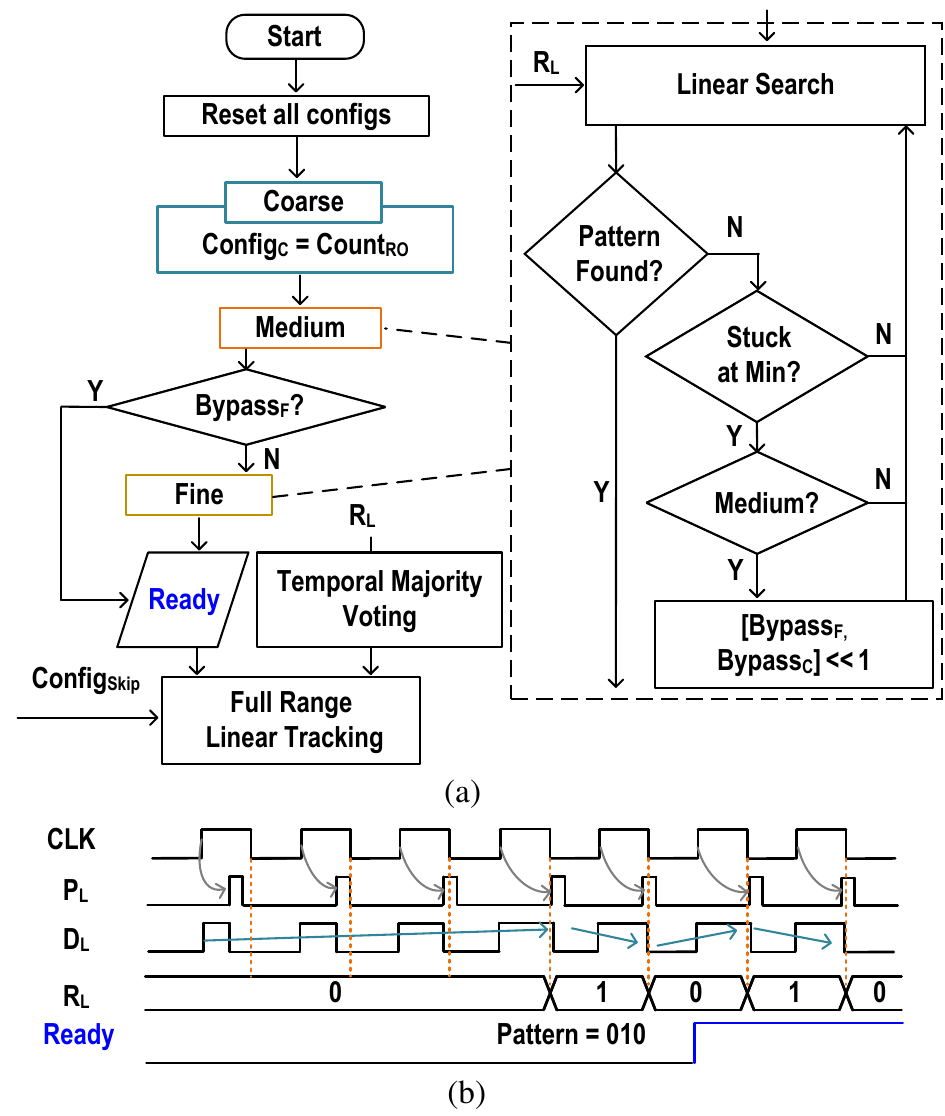}
  \vskip -2ex
  \caption{(a) The FSM and (b) the locking process to lock to the clock signal.}
  \label{fsm}
\end{figure}

Real-time delay-locking is implemented on-chip as a Finite State Machine (FSM) to eliminate off-chip calibration. The FSM automatically configures the delay line to lock to the pulse width of the clock. As shown in Fig.~\ref{fsm}, it determines the configurations for the three delay stages step by step. The FSM starts with all configurations reset to zero and then sequentially calibrates coarse, medium, and fine stages. The coarse stage tuning can be completed in one cycle: the FSM times the system clock's pulse width with the ring oscillator inside the coarse stage. The configuration for the coarse stage is exactly the ring oscillator cycles it takes for the system clock to rise and fall back to zero again. 

To tune the medium or the fine stage, the FSM linearly searches for the delay configuration. The FSM increments the delay control bits by one until $R_L$ becomes one, which indicates that the pulse width of the replica is longer than that of the system clock. Then, the FSM will decrease the control bits to reduce the delay. At this point, the locked delay is the maximum delay that is less than the clock's pulse width. Therefore, the linear search is completed if the pattern ``010'' is found on $R_L$. On the other hand, if the minimum configuration (i.e., all zeros) is already too large, the FSM will sequentially try bypassing the coarse and fine stages to reduce the delay offset of the configurable delay line. If the FSM still cannot lock to the clock, it terminates and reports an error. 

\subsection{Real-Timing Delay Tuning under Frequency Drift}

\begin{figure}[t]
  \centering
  \includegraphics[width=\linewidth]{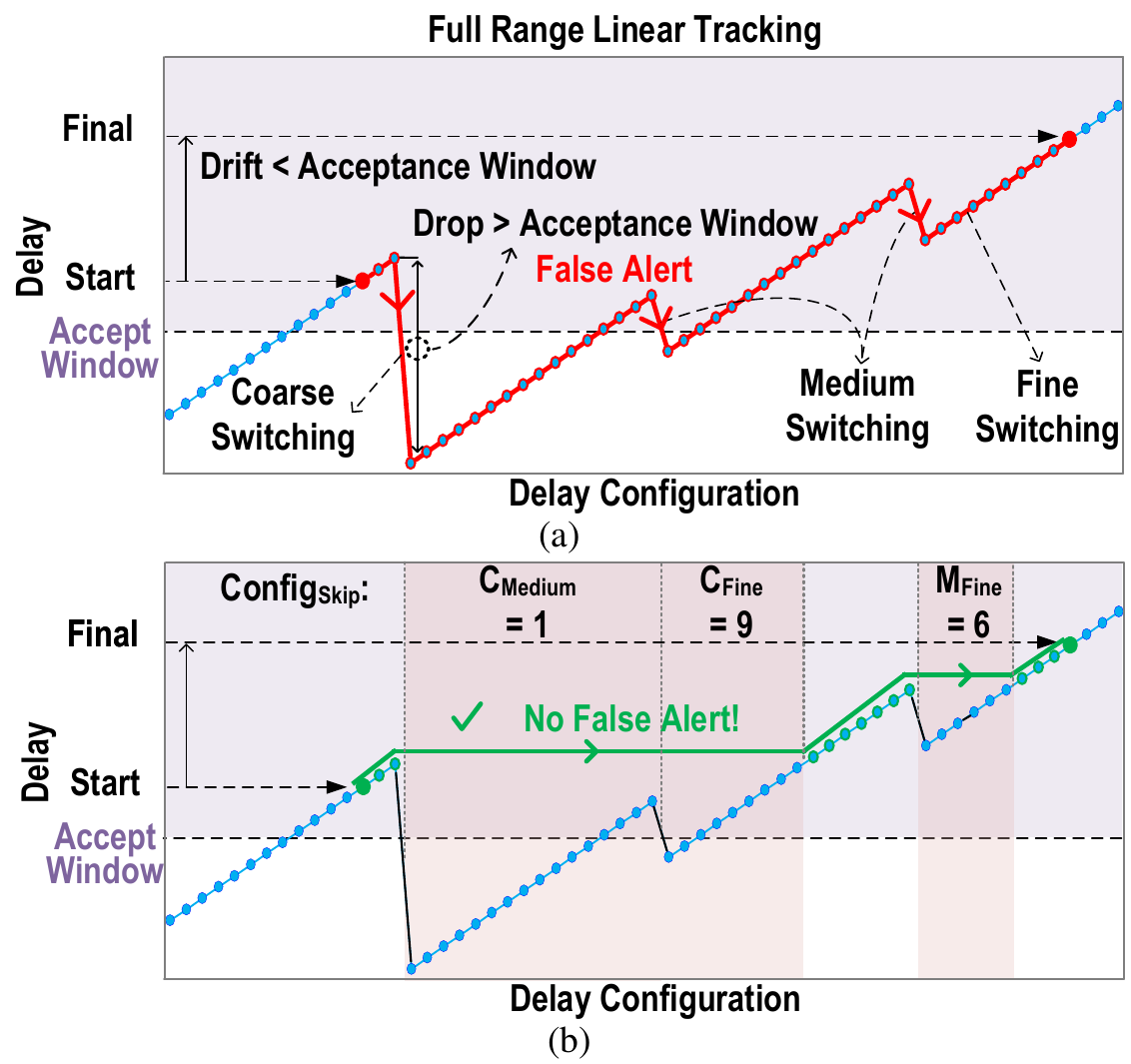}
  \vskip -2ex
  \caption{(a) Overlapping between the delay stages leads to delay drops when switching. A large drop can set the drop outside the acceptance window and trigger a false alert when the FSM is tuning the delay line. (b) Programmable configuration skips are introduced to ensure that the delay increases monotonically with the configuration code.}
  \label{config_skip}
\end{figure}

The initial locking process above is performed once right after the system powers up or resets. The monitor starts FIA detection after this initial phase. The FSM then switches into the Full Range Linear Tracking mode. In this mode, the FSM dynamically matches the delay configurations to the clock signal using the finest step. Temporal Majority Voting is used to low-pass filter $R_L$ for better stability. The purpose of this mode is to tolerate and track the slow drifting of the clock due to clock source drifts, environmental changes, or device aging. Otherwise, the clock signal may drift out of the acceptance window without any FIAs, which will trigger false alerts. 

Clock drift is not the only source of potential false alerts. 
The coarse, medium, and fine stages in the fully synthesized delay line will not match perfectly to have a monotonic relation with the digital delay calibration bits. This is an intentional design choice because the automatic placement and routing procedure inevitably introduces delay mismatches unless significant efforts are made to match the layout. Indeed, the tuning range of a finer stage is designed to be greater than one step in the coarser stage. As shown in Fig.~\ref{config_skip}a, when the finer tuning stage reaches the maximum value and returns to zero, the coarser stage will increase by one. Because of the mismatch, the delay will decrease even if the overall configuration bits are increasing. This delay drop could cause the clock edge to fall out of the tolerance window and trigger a false alert. 
To overcome this issue, programmable configuration skips are introduced to ensure that the delay monotonously increases as the configuration bits increase. Fig.~\ref{config_skip}b demonstrates one example in which one medium step and nine fine steps are skipped after each coarse step increase while six fine steps are skipped for each medium step increase. Specifically, when the coarse configuration increments by one, the medium bits and the fine bits start from one and nine, respectively, instead of zero. This scheme guarantees that false alerts are minimized even with the presence of mismatches between the delay tuning ranges. 

\section{Automatic Generation of the FIA Monitor}
\label{gen_framework}

\begin{figure}[t]
  \centering
  \includegraphics[width=\linewidth]{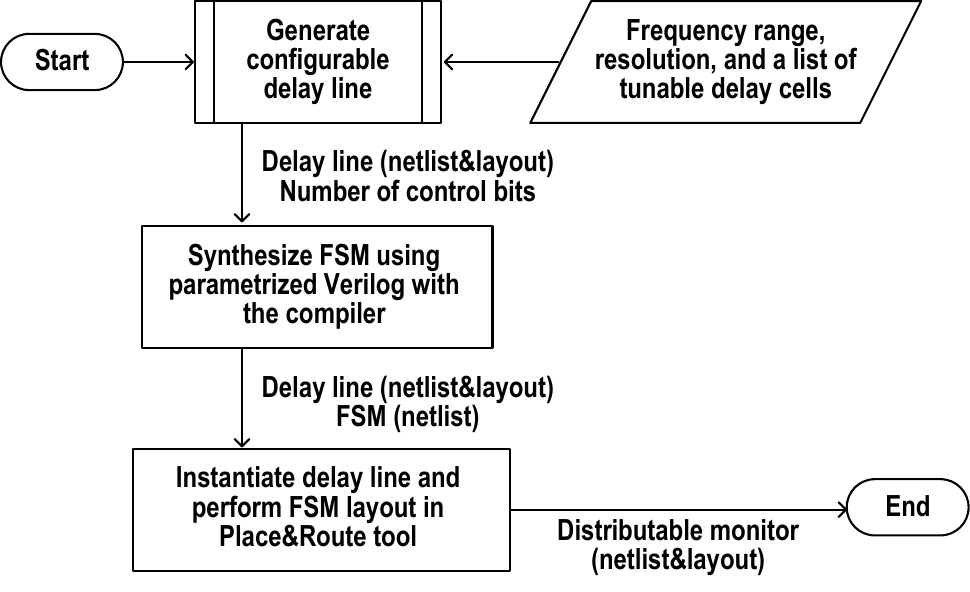}
  \vskip -2ex
  \caption{The end-to-end automated workflow to generate the FIA monitor.}
  \label{overall_flow_chart}
\end{figure}

To facilitate the agile implementation of our FIA monitor in different systems, we develop an end-to-end design framework to fully automate the design and implementation process (Fig.~\ref{overall_flow_chart}). Since the overall monitor design is fully synthesizable without any manual constraints, only the configurable delay line needs to be redesigned for different systems and process nodes to achieve optimal performance. 
The framework takes a list of delay cells to use, the desired locking frequency range, and the target resolution as inputs. It also needs access to the standard cell library as well as the standard EDA tools in the digital design flow. It first generates the configurable delay line, and then combines it with the tuning FSM compiled from Verilog code. The framework is optimized to produce the circuit with the minimum area that satisfies the frequency range and resolution demands. We build the framework using Python, which interfaces with EDA tools with Tcl scripts.

\subsection{Configurable Delay Line Optimization} 

\begin{figure}[t]
  \centering
  \includegraphics[width=\linewidth]{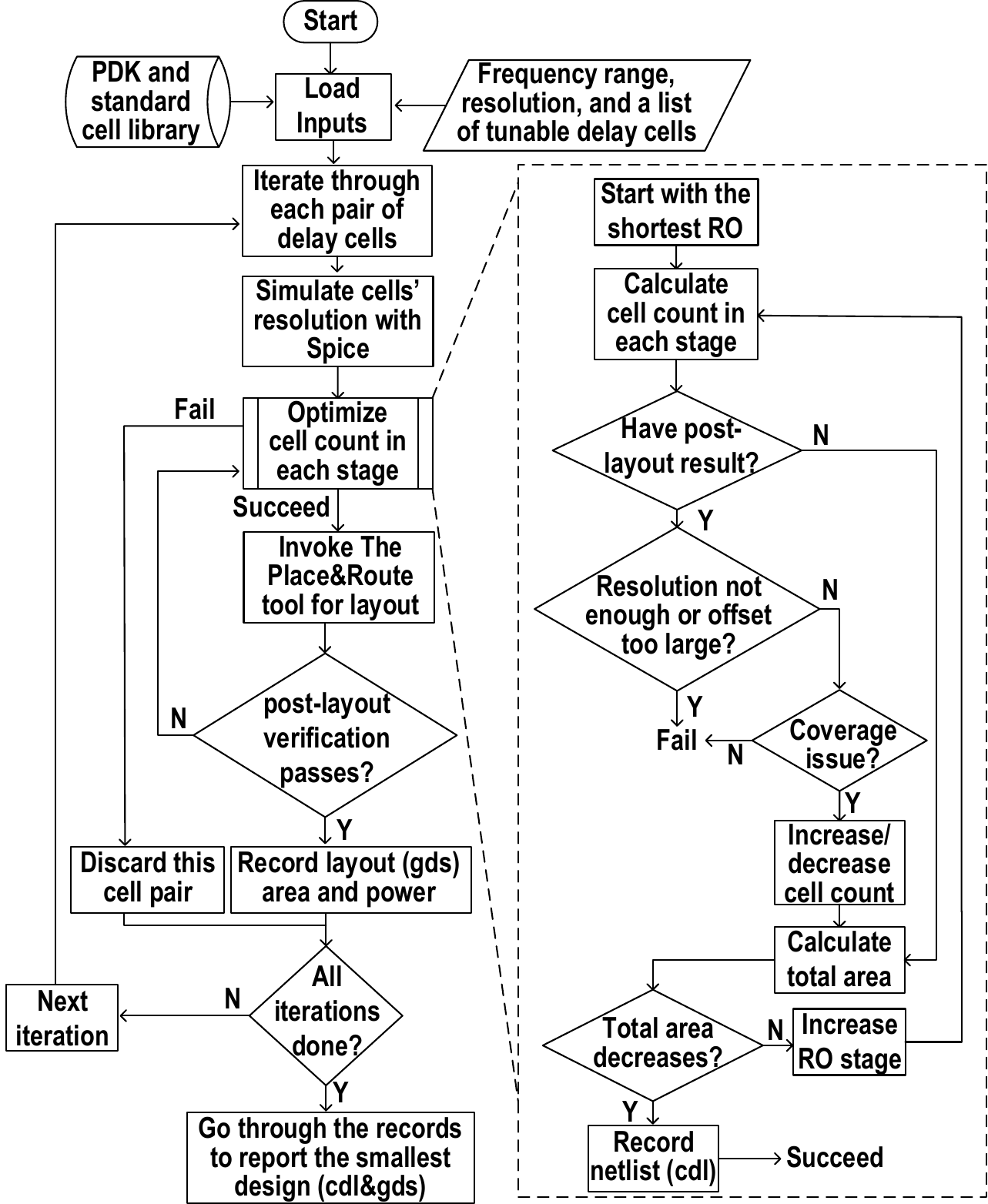}
  \vskip -2ex
  \caption{The flow chart for automating the configurable delay line design.}
  \label{cdl_flow_chart}
\end{figure}

The workflow for designing the delay line is summarized in Fig.~\ref{cdl_flow_chart}. The framework needs to decide which delay cells to use for the medium and the fine stages, as well as how many of these cells to use. The coarse stage is fixed to be the RO-counting topology because this topology's area scales sublinearly with the delay tuning range. The delay line generator iterates through all pairs of medium and fine cells in the user-specified delay cell list. 
For each pair of cells, Python generates an evaluation netlist and invokes the spice simulator to determine the cell's delay when the control bit is 0 and 1.
The resolution of the cell is calculated as the delay difference when the cell is turned on and off. The cell with the greater delay difference is assigned as the medium stage, with the other as the fine stage. Since the delay tuning range of the finer stage must be greater than one delay tuning step in the coarse stage, the number of cells to use in each stage is straightforward: the ratio between the two stages' resolution is rounded up to be the number of cells in the finer stage. The user may specify an additional margin to ensure that the finer stage's range is enough considering the variations. If a $50\%$ margin is used, the number of cells in the finer stage becomes 1.5 times the resolution of the coarser stage divided by the finer stage's resolution, rounded up to the nearest integer. 

\begin{figure}[t]
  \centering
  \includegraphics[width=0.85\linewidth]{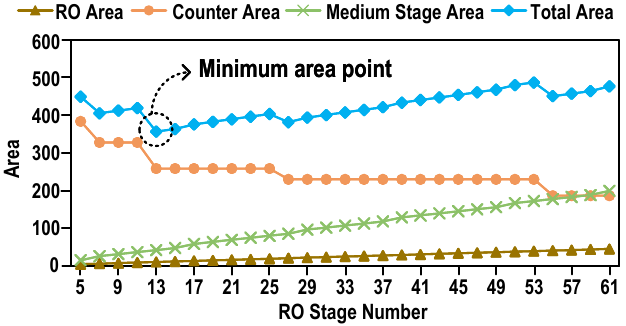}
  \vskip -2ex
  \caption{The delay line area breakdown.}
  \label{delay_area_breakdown}
\end{figure}

While the coarse stage's topology is fixed, the framework still has to optimize the number of stages in RO. A shorter RO leads to a smaller coarse step, which reduces the number of medium cells. However, since the delay line needs to cover the user-specified delay tuning range, a smaller coarse step means that a larger counter must be used. To optimize under this tradeoff, the framework first simulates the period of the RO with the smallest number of stages. This period is also the smallest coarse step, which determines the maximum number of counter bits. Then, the framework increases the number of stages in the RO by two at a time, calculates the number of cells in each stage, and evaluates the total area of the delay line. As shown in Fig.~\ref{delay_area_breakdown}, the counter's footprint dominates the total area in the beginning. As the number of stages in the RO increases, the counter becomes smaller, and the medium stage eventually dominates the total area. 

Once the framework determines the RO design, it performs post-layout simulations to evaluate the current design built from the two selected delay cells. 
It executes standard EDA tools with Tcl scripts to place and route the delay line and obtains the post-layout netlist. Spice simulations are run on the post-layout netlist by Python to extract the delay tuning range and the resolution of each stage.
The framework first verifies that the tuning range of the finer stage indeed covers one tuning step of the coarser stage and then examines whether the resolution and the tuning range pass the user-specified requirements. If all constraints are met, the framework records the area and optionally power data of the current topology. Otherwise, the framework tries to fine-tune the design and discard the current circuit if the fine-tuning attempt fails. In either case, the framework finishes evaluating the selected pair of delay cells and moves to the next. 

After exhaustively searching for all designs that meet the specifications, the framework decides the one with the smallest area. The framework also supports optimization based on power consumption when the monitor's power is a concern. To enable this function, the user must define the frequency at which the monitor's power should be evaluated. The framework can combine the area and the power consumption with programmable weights as the cost function for optimization. 

\subsection{FSM Integration}
After obtaining the optimized configurable delay line, the framework attaches the auto-calibration FSM to it and completes the monitor design. The number of control bits in each stage is parameterized so the framework can reuse the FSM for any delay line design without modifying the code. From this step, the framework follows the standard backend flow to generate the monitor layout. The user can then instantiate the monitor design wherever FIA protection is desired. 

\section{Measurement Results}
\label{measure_results}

\subsection{Automated Monitor Design and Silicon Prototype}

\begin{table*}[t]
\caption{\textbf{Monitors Designed by the Automation Framework}}
\centering
\setlength{\tabcolsep}{3.6pt}
\vskip -1.5ex
\begin{tabular}{|c|cc|ccc|cccc|}
\hline
\multicolumn{3}{|c|}{\textbf{User Inputs}} & \multicolumn{3}{c|}{\textbf{Generated Structure}} & \multicolumn{4}{c|}{\textbf{Post-Layout Evaluation}} 
\\\hline
Technology & \multicolumn{1}{c|}{Frequency Range} & \parbox[c][0.8cm]{45pt}{\centering{Resolution\\Requirement}} & \multicolumn{1}{c|}{\parbox[c][0.8cm]{38pt}{\centering{RO Stage\\Number}}} & \multicolumn{1}{c|}{\parbox[c][0.8cm]{40pt}{\centering{Medium\\Bit Number}}} & Fine Stage & \multicolumn{1}{c|}{\parbox[c][0.8cm]{70pt}{\centering{Frequency Range\\w/o Resolution Loss}}} & \multicolumn{1}{c|}{\parbox[c][0.8cm]{48pt}{\centering{Frequency\\w/ Bypassing}}} & \multicolumn{1}{c|}{Resolution} & \parbox[c][0.8cm]{21pt}{\centering{Area\\($\mu$m$^2$)}} 
\\\hline
65nm & \multicolumn{1}{c|}{2MHz - 600MHz} & 10ps & \multicolumn{1}{c|}{13} & \multicolumn{1}{c|}{8b} & 4b NAND2 & \multicolumn{1}{c|}{1.84MHz - 698MHz} & \multicolumn{1}{c|}{1.13GHz} & \multicolumn{1}{c|}{6.84ps} & 1500 
\\\hline
65nm & \multicolumn{1}{c|}{2MHz - 700MHz} & 20ps & \multicolumn{1}{c|}{13} & \multicolumn{1}{c|}{8b} & 3b NOR2 & \multicolumn{1}{c|}{1.89MHz - 768MHz} & \multicolumn{1}{c|}{1.18GHz} & \multicolumn{1}{c|}{17.1ps} & 1464 
\\\hline
65nm & \multicolumn{1}{c|}{500kHz - 10MHz} & 100ps & \multicolumn{1}{c|}{17} & \multicolumn{1}{c|}{11b} & N/A & \multicolumn{1}{c|}{348kHz - 1.09GHz} & \multicolumn{1}{c|}{1.09GHz} & \multicolumn{1}{c|}{69.9ps} & 1596 
\\\hline
65nm & \multicolumn{1}{c|}{500MHz - 600MHz} & 10ps & \multicolumn{1}{c|}{0} & \multicolumn{1}{c|}{9b} & 4b NAND2 & \multicolumn{1}{c|}{369MHz - 704MHz} & \multicolumn{1}{c|}{1.16GHz} & \multicolumn{1}{c|}{7.44ps} & 1188 
\\\hline
28nm & \multicolumn{1}{c|}{2MHz - 2GHz} & 2ps & \multicolumn{1}{c|}{{21}} & \multicolumn{1}{c|}{{12b}} & {4b NAND2} & \multicolumn{1}{c|}{{1.86MHz - 2.07GHz}} & \multicolumn{1}{c|}{{2.66GHz}} & \multicolumn{1}{c|}{{1.92ps}} & {459}
\\\hline
{28nm} & \multicolumn{1}{c|}{{1.5GHz - 2.5GHz}} & {5ps} & \multicolumn{1}{c|}{{0}} & \multicolumn{1}{c|}{{16b}} & {2b NOR3} & \multicolumn{1}{c|}{{1.43GHz - 2.66GHz}} & \multicolumn{1}{c|}{{3.07GHz}} & \multicolumn{1}{c|}{{4.92ps}} & {374}
\\\hline
\end{tabular}
\label{gen_result}
\end{table*}

\begin{figure}[t]
  \centering
  \includegraphics[width=\linewidth]{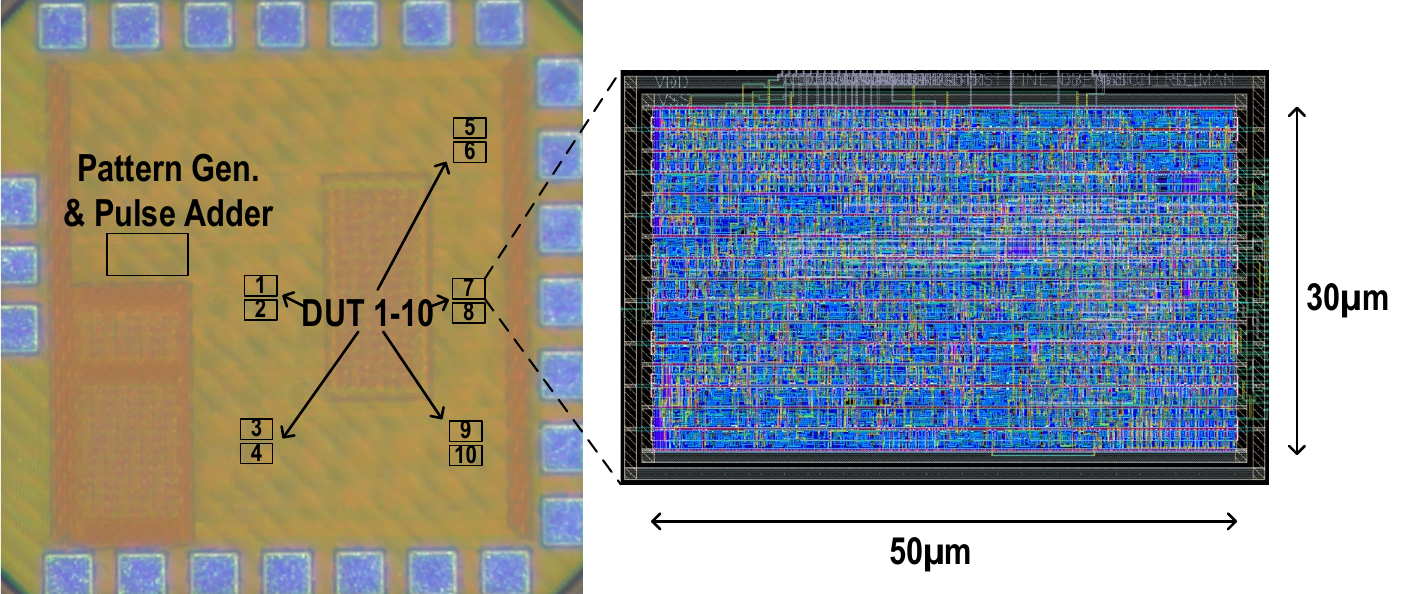}
  \vskip -2ex
  \caption{The 65nm test chip and the automatic P\&R layout of one unit.}
  \label{die_photo}
\end{figure}

We evaluated the automated framework with four specifications in 65nm CMOS, shown in the first four rows of Table~\ref{gen_result}. {To demonstrate technology scalability, we also generated two designs with a 28nm CMOS PDK, with more details in Section~\ref{tech_scale}.} The first row is a scenario demanding both a wide frequency range and a high resolution. For comparison, the second row showcases that by lowering the resolution requirement, higher frequencies can be achieved without bypassing the fine stage. The design is also slightly smaller than the first scenario. The third and fourth rows demonstrate the use cases for low and high frequencies, respectively. The low-frequency scenario does not require the fine stage, while the counter-based coarse stage is omitted when only high frequencies are targeted. In all scenarios, the path-selection-based delay line is chosen as the medium stage because of its small overhead in terms of delay offset and footprint. 

The design shown in the first row of Table~\ref{gen_result} is fabricated and verified in 65nm LP CMOS. The design is automatically placed and routed by industry-standard software without any manual adjustment or optimization. Each chip contains ten monitors, and each monitor takes a footprint of $30\mu m\times 50\mu m$ (Fig.~\ref{die_photo}), or equivalently $355,000F^2$ {if calculated with the feature size (F) of 65nm}. 

\begin{figure}[t]
  \centering
  \includegraphics[width=\linewidth]{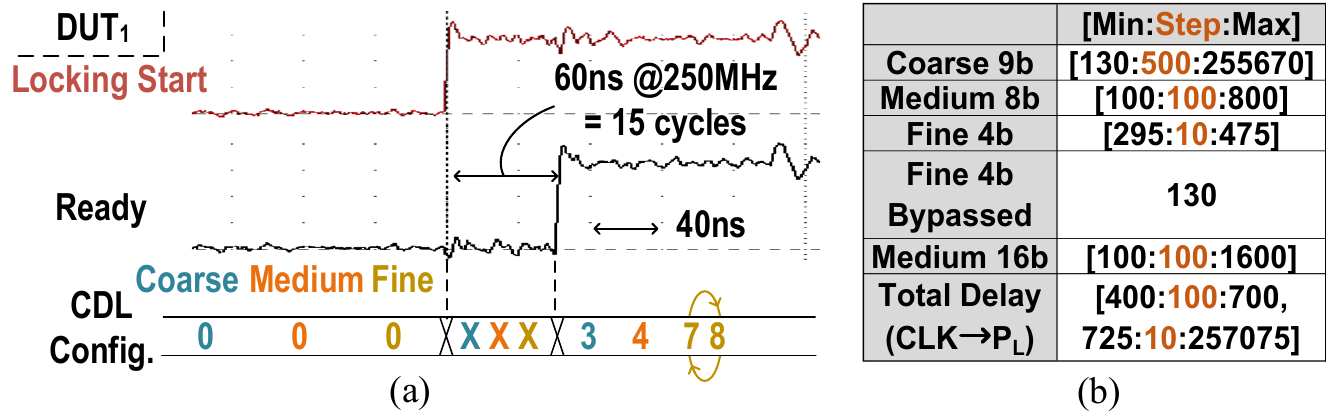}
  \vskip -2ex
  \caption{The locking process is shown in (a) and the delay range and resolution for which stage are shown in (b). The measurement resolution is limited by the jitter performance of the clock source}
  \label{delay_line_test}
\end{figure}

\begin{figure}[t!]
  \centering
  \includegraphics[width=\linewidth]{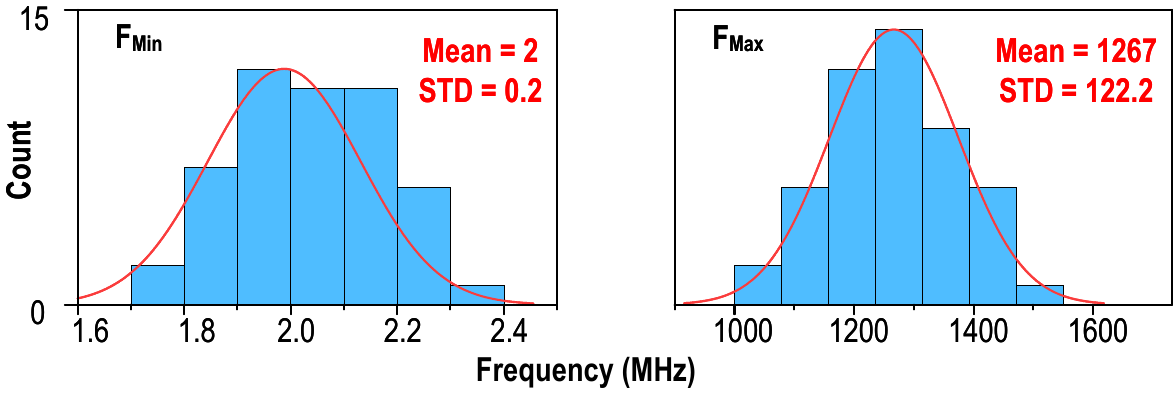}
  \vskip -2ex
  \caption{The locking frequencies across 50 DUTs at 25$^\circ C$ and $1.2V$.}
  \label{freq_var_p}
\end{figure}

\subsection{Evaluations of the Synthesized Delay Line}

Fig.~\ref{delay_line_test}a showcases the locking process of the configurable delay line. In this particular case, the clock frequency is $250MHz$, and the locking process takes 15 cycles, or $60ns$. Depending on the pulse width of the clock, it takes from 7 to 26 cycles for the monitor to lock to the clock: 1 cycle for the coarse stage, 3 to 9 cycles for the medium stage, and 3 to 17 cycles for the fine stage. The lower bound of the medium and the fine stage is determined by the length of the ``010'' pattern, and the number of possible configurations in each stage decides the upper bound. 

Fig.~\ref{delay_line_test}b summarizes the range and the resolution of the three delay tuning stages. The tuning step is extrapolated from the maximum and the minimum delay for each stage, assuming that the delay increase is linear. While the fine stage has the finest resolution, the offset of the fine stage is the greatest. Therefore, the option to bypass the fine stage is implemented. With this option, the total tuning range of the configurable delay line is $400ps$ to $257ns$ at $1.2V$, or equivalently $2MHz$ to $1.26GHz$. The lower frequency limit can be reduced by increasing the number of configuration bits in the coarse stage. As the coarse stage is based on counting the ring oscillator cycles, this change incurs almost no overhead. 

\begin{figure}[t]
  \centering
  \includegraphics[width=\linewidth]{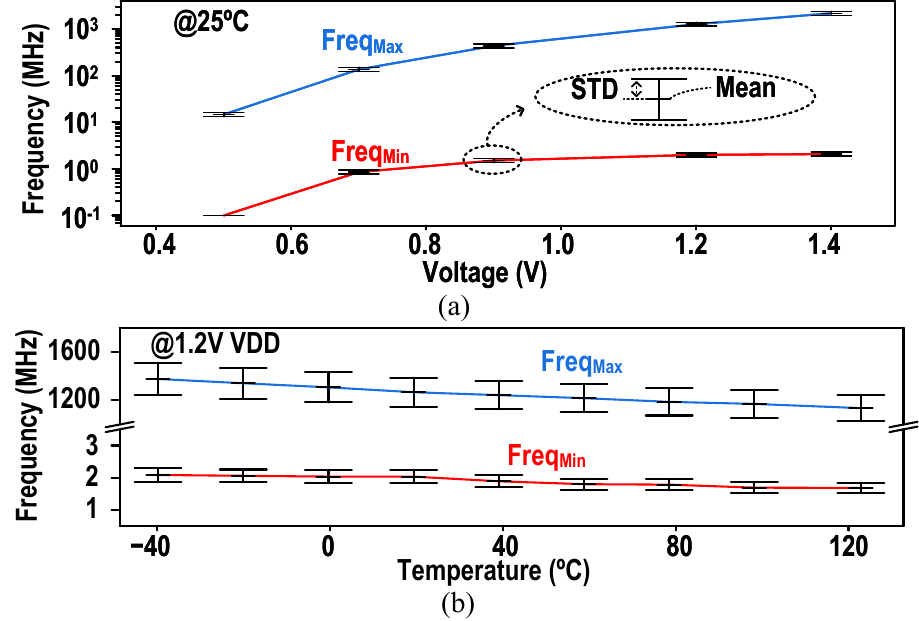}
  \vskip -2ex
  \caption{Locking frequency variations across (a) voltage and (b) temperature.}
  \label{freq_var_vt}
\end{figure}

In total, fifty DUTs are tested, and the locking frequency results are shown in Fig.~\ref{freq_var_p} and Fig.~\ref{freq_var_vt}. The locking frequency range shows minimum variations across fifty DUTs. The temperature also does not significantly affect the frequency range. The locking frequency does change with the supply voltage. This dependency is anticipated as the monitor is expected to reflect the logic gates' delay at different voltages. 

\subsection{Clock Glitching Attacks}

\begin{figure}[t]
  \centering
  \includegraphics[width=\linewidth]{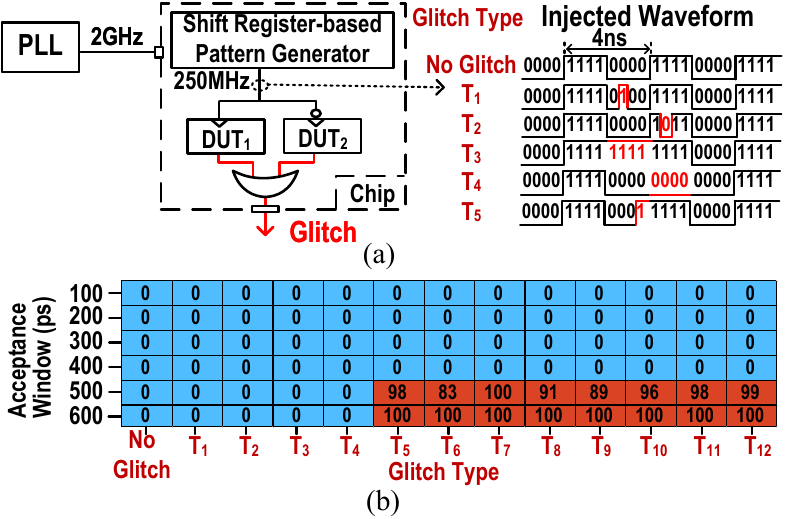}
  \vskip -2ex
  \caption{(a) The diagram for the on-chip pattern generator and (b) the number of missed alerts for 100 trials with each configuration.}
  \label{clock_glitch_miss_rate}
\end{figure}

\begin{figure}[t]
  \centering
  \includegraphics[width=\linewidth]{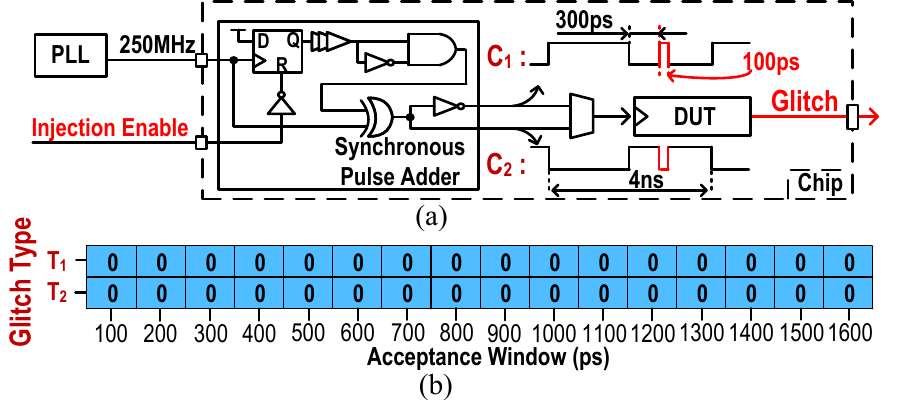}
  \vskip -2ex
  \caption{(a) The fast pulse addition circuit. (b) $\text{T}_\text{1-2}$ types of clock glitches can always be detected.}
  \label{pulse_add}
\end{figure}

The miss rates of the monitor for each clock glitch type with different acceptance window configurations are listed in Fig.~\ref{clock_glitch_miss_rate}b. Clock glitches of the 12 types are injected into the normal $250MHz$ clock using the on-chip $2GHz$ arbitrary pattern generator. The glitch width is set to be $500ps$. A conceptual diagram for this testing setup is shown in Fig.~\ref{clock_glitch_miss_rate}a. A pair of monitors {is} tested in order to detect every type of clock glitch, as explained in Section \ref{FIA_monitor}. When the acceptance window is less than or equal to $400ps$, all 12 types of clock glitches can be accurately detected. At $500ps$ or larger, the monitor cannot distinguish clock glitches of type $\text{T}_\text{5-12}$ from clock jitters. $\text{T}_\text{1-4}$, on the other hand, can still be detected by the monitor. $\text{T}_\text{3-4}$ merges three clock pulses into a longer clock phase. These glitches can always be identified because the acceptance window will never exceed one pulse width of the clock. $\text{T}_\text{1-2}$ involves splitting a clock pulse into three segments. Such changes can be detected as long as any of the modified clock edges falls out of the acceptance window. 

To better demonstrate the monitor's remarkable ability to detect pulse injection attacks on the clock, a fast pulse addition circuit is implemented on the chip. Fig.~\ref{pulse_add}a depicts the principle of the adder circuit and how it interacts with the monitor. This testing circuit can inject pulses as short as $100ps$ into the clock signal. Again, our monitor can detect all of these glitches even when the acceptance window is greater than the injected pulse width, as shown in Fig.~\ref{pulse_add}b. 

The testing results prove that our monitor can detect all 12 types of glitches, provided that the acceptance window setting is reasonable. Since real-world clock glitch attacks can always be viewed as a combination of these 12 types, our monitor is able to detect FIAs based on clock glitches precisely. 

\subsection{Fault Injections via Voltage, EM, and Temperature}
Besides the clock signal, FIAs can also be mounted by altering the delay of the logic gates. These alternations can be achieved by manipulating the operating environment of the chip, such as the supply voltage, the temperature, or even EM signals. The monitor is evaluated to determine if it can detect these physical fault injection attacks. The clock frequency is $100MHz$ in all of the tests below. 

\begin{figure}[t]
  \centering
  \includegraphics[width=\linewidth]{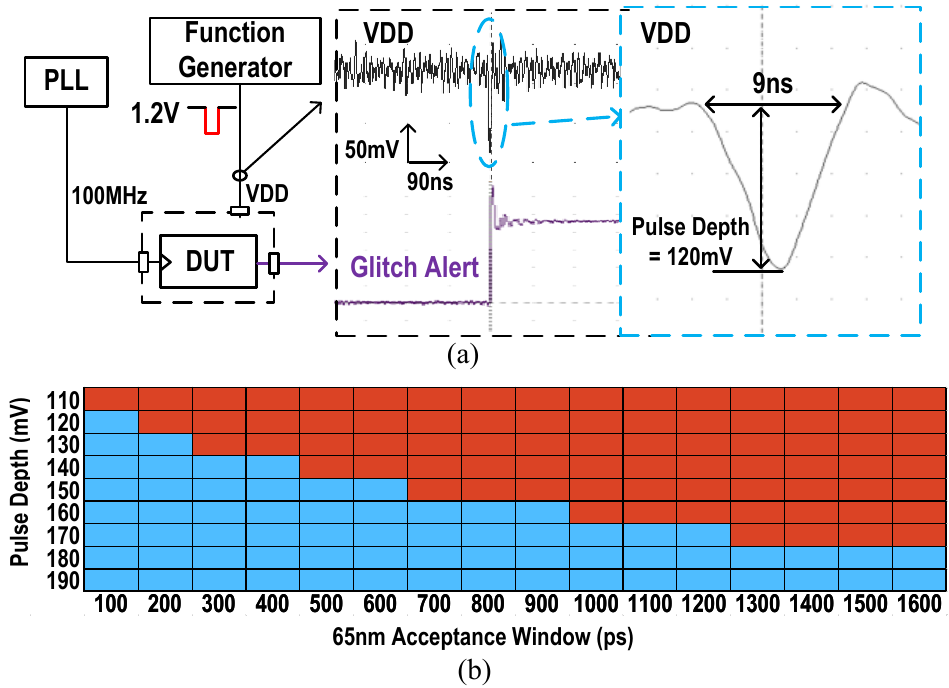}
  \vskip -2ex
  \caption{(a) Testing setup for the voltage glitch attack. The EMP attack would work in a similar fashion. (b) A grid is marked blue if the glitch can be detected at the corresponding operation point. By shrinking the acceptance window, more accurate detection can be achieved.}
  \label{test_v_glitch}
\end{figure}

\subsubsection{Voltage Attack}
An off-chip function generator is used to inject glitches into the power supply of the circuit. The monitor can correctly identify voltage glitches as small as $120mV$ whose duration is $9ns$. By configuring the acceptance window, different sensitivity levels can be achieved, as shown in Fig.~\ref{test_v_glitch}. Note that the supply voltage is quite noisy in the testing setup. False alerts are avoided by choosing the appropriate acceptance window size. 

\begin{figure}[t]
  \centering
  \includegraphics[width=\linewidth]{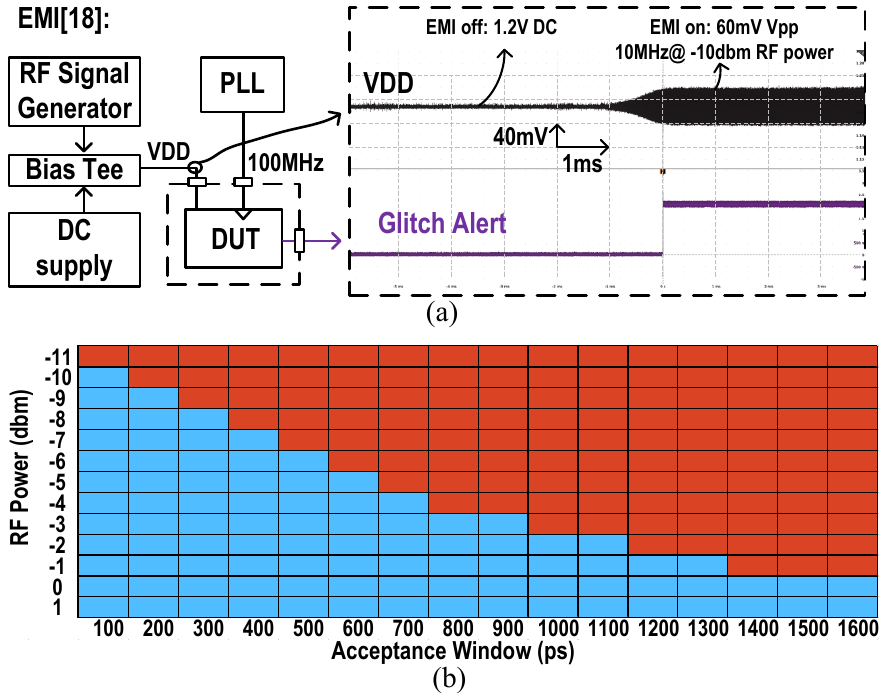}
  \vskip -2ex
  \caption{(a) EM attack test setup. (b) Success/fail summary at different RF power and acceptance window settings.}
  \label{test_em}
\end{figure}

\begin{figure}[t]
  \centering
  \includegraphics[width=\linewidth]{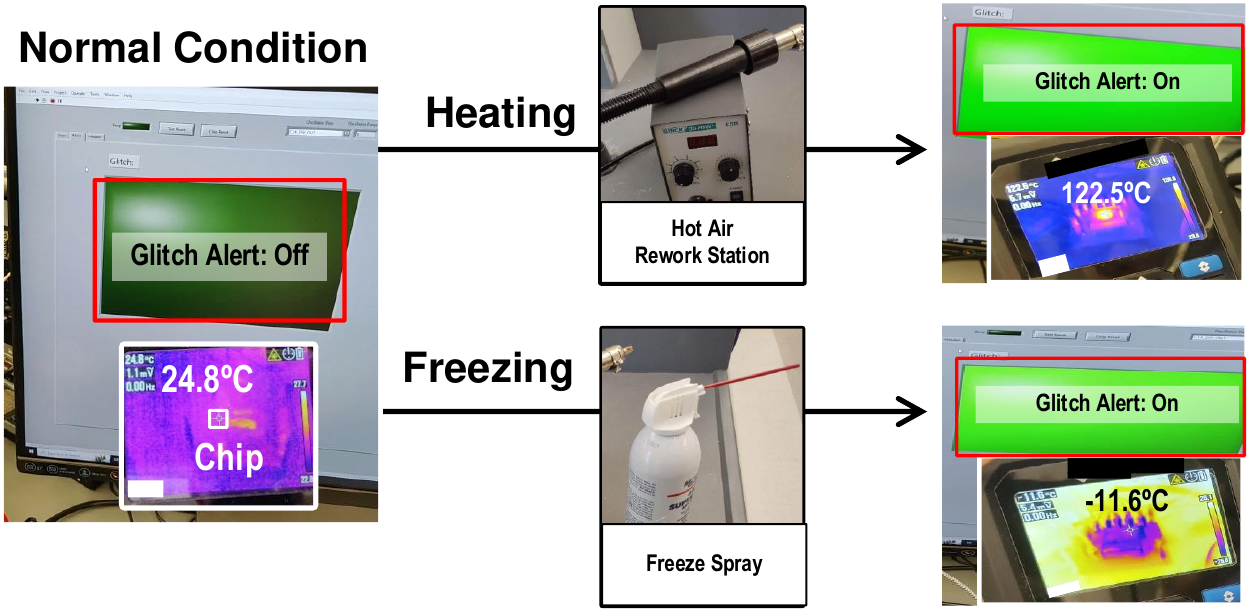}
  \vskip -2ex
  \caption{The temperature attack testing procedures.}
  \label{test_temp}
\end{figure}

\begin{table}[t!]
\caption{\textbf{Temperature Attack Parameters}}
\centering
\setlength{\tabcolsep}{3.6pt}
\renewcommand{\arraystretch}{1.2}
\vskip -1.5ex
\begin{tabular}{|c|c|c|c|}
\hline
& Heating Attack & Freezing Attack & Temperature Drift
\\\hline
Temperature ($^\circ$C) & $25\rightarrow122$ & $25\rightarrow-11$ & $-40\rightarrow125$
\\\hline
\parbox[c][0.8cm]{50pt} {\centering{Average Slew\\Rate ($^\circ$C/min)}} & 1200 & -600 & 2
\\\hline
\parbox[c][0.8cm]{50pt}{\centering{Testing\\Equipment}} & \parbox[c][0.8cm]{50pt}{\centering{Hot air\\rework station}} & Freeze spray & \parbox[c][0.8cm]{50pt}{\centering{Temperature\\chamber}}
\\\hline
Monitor Result & Glitch detected & Glitch detected & No glitch
\\\hline
\end{tabular}
\label{temp_attack}
\end{table}

\subsubsection{EM Attack}
Similar to voltage attack testing, an RF signal is generated off-chip and coupled to the chip's power supply. EMP attacks are not evaluated due to equipment restrictions, but EMP attacks have similar, if not weaker, effects as the voltage glitch attack. Fig.~\ref{test_em}a shows the testing setup for EMI attacks. The RF signal injects into the supply voltage a $10MHz$ noise with $>-10\text{dBm}$ power, the same EM frequency as used in \cite{fujimoto_detection_2018}. This supply disturbance has an equivalent voltage of {$60mV_{pp}$}, which is smaller than the $90mV$ interference in that work. Our monitor successfully detects this more subtle EM FIA, as shown in Fig.~\ref{test_em}b. The acceptance window can also be configured to tune the detection threshold of the injected noise. 

\subsubsection{Temperature Attack}
Freeze spray is applied to the chip to quickly cool down the circuit, while a hot air rework station is used to ramp up the temperature in a short time (Fig.~\ref{test_temp}). As summarized in Table~\ref{temp_attack}, the temperature slew rate for the cooling (heating) process is $-600^\circ C/min$ ($1200^\circ C/min$). The monitor detects the injection attack in both cases. On the contrary, the chip is placed in a temperature chamber to simulate normal environmental temperature change. With the help of the real-time delay tuning functionality, the monitor does not raise a false alert in this case. 

\begin{figure}[t]
  \centering
  \includegraphics[width=\linewidth]{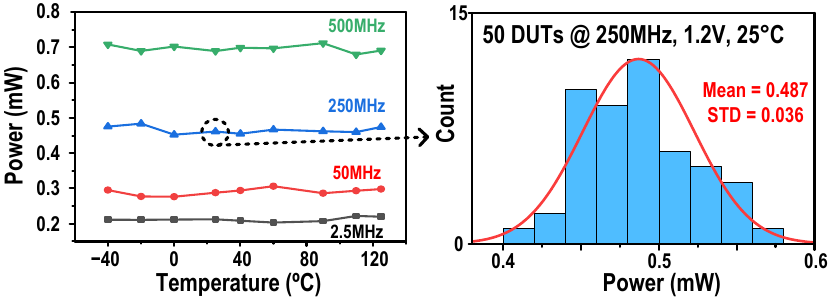}
  \vskip -2ex
  \caption{DUT power across temperature and variations in power across DUTs.}
  \label{power_t}
\end{figure}

\begin{figure}[t]
  \centering
  \includegraphics[width=\linewidth]{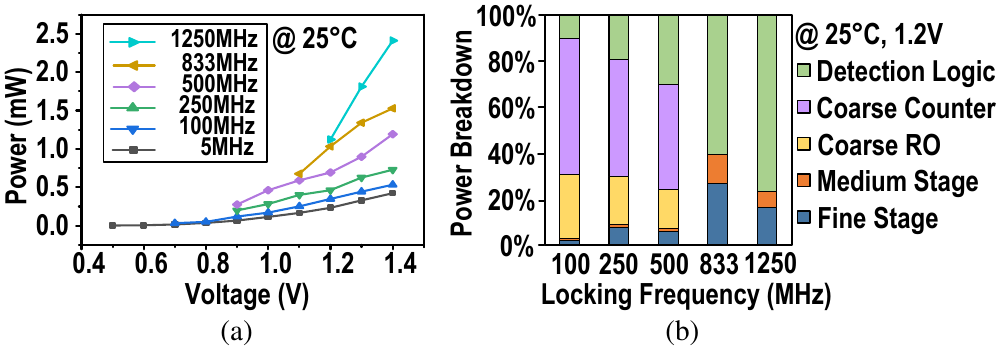}
  \vskip -2ex
  \caption{(a) Power consumption across voltage. (b) Specific power breakdown. Starting from $833MHz$, the coarse stage is bypassed.}
  \label{power_v}
\end{figure}

\begin{table*}[t!]
\caption{\textbf{Comparison Table with State-of-the-Art FIA Monitors}}
\centering
\setlength{\tabcolsep}{3.6pt}
\renewcommand{\arraystretch}{1.2}
\vskip -1.5ex
\begin{tabular}{|c|c|c|c|}
\hline
&
\textbf{This Work} & ISSCC'23 \cite{kumar_100gbps_2023} & VLSI'22 \cite{song_fll-based_2022}
\\\hline
Technology & \textbf{65nm} & 4nm & 5nm
\\\hline
Protection Target & \textbf{Design Agnostic} & AES-256 & Design Agnostic
\\\hline
Principle & \textbf{Pulse Width Comparison} & Error Checking & High-Frequency Sampling
\\\hline
Digital Design & \textbf{Fully Synthesizable} & Fully Synthesizable & Partially Digital
\\\hline
Voltage (V) & \textbf{0.5 - 1.4} & 0.75 & 0.5 - 1.0
\\\hline
Temperature ($^\circ$C) & \textbf{-40 - 125} & 25 & 25
\\\hline
Power (mW) & \textbf{0.487}$^\text{a}$ & - & 0.8025$^\text{b}$
\\\hline
Area (MF$^2$) & \textbf{0.355} & 244.56 & 192
\\\hline
Monitor Precision & \textbf{DLL Delay Step} & - & FLL Period
\\\hline
Clock Frequency & \textbf{2MHz - 1.26GHz} & 0 - 780MHz & 0 - 40MHz
\\\hline
Target Attacks & \textbf{Clock, Voltage, EM, Temperature Attacks} & Any Fault Attack on AES & Low-Frequency Clock Glitches
\\\hline
\multicolumn{2}{l}{a: measured @ 1.2V VDD, locking to 250MHz clock.} & \multicolumn{2}{l}{b: measured @ 0.75V VDD, locking to 40MHz clock.}

\end{tabular}

\label{comp}
\end{table*}

\subsection{Power Consumption and Breakdown}
The monitor is tested across a wide range of temperatures and voltages. The power consumption at each tested point is depicted in Fig.~\ref{power_t} and Fig.~\ref{power_v}. The monitor is able to function across the automotive temperature range, which is $-40$ to $125^\circ C$. It can be inferred from the figure that the tuning frequency is insensitive to the environment temperature. The monitor is also validated under a versatile of voltages, from $0.5V$ to $1.4V$. The average power of the fifty DUTs is $0.487mW$ at $1.2V$, $25^\circ C$, when they are locked to a $250MHz$ clock. Fig.~\ref{power_v}b shows the power breakdown at various locking frequencies. The coarse stage dominates the power consumption at lower frequencies because it contains a counter triggered by an active RO. After the coarse stage is bypassed at $833MHz$, the delay window logic consumes the majority of the power.

\begin{figure}[t]
  \centering
  \includegraphics[width=\linewidth]{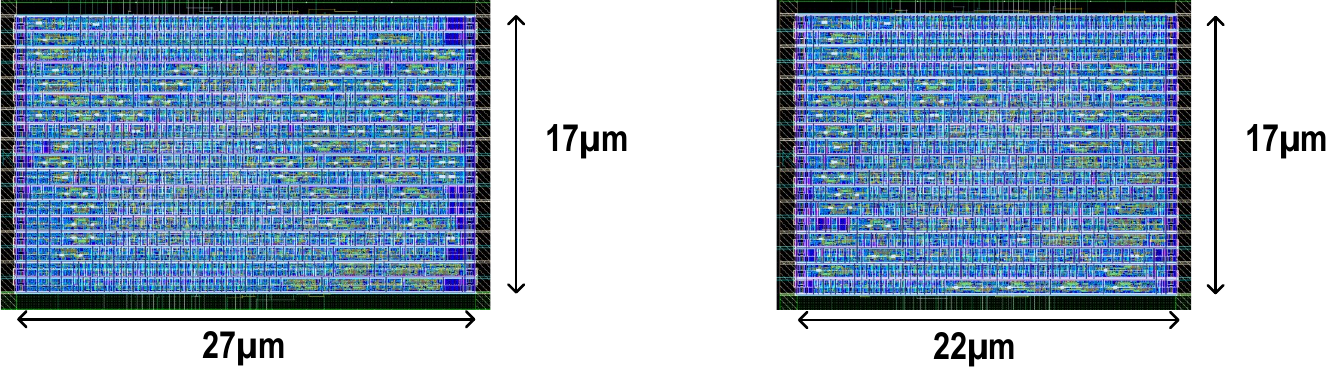}
  \vskip -2ex
  \caption{Generated 28nm layout for the fifth row and the sixth row in Table~\ref{gen_result}.}
  \label{28_apr}
\end{figure}

\section{Technology Scalability of the Framework}
\label{tech_scale}

Table~\ref{gen_result} showcases two sample FIA monitor designs automatically generated by our framework in 28nm CMOS. 
The fifth row represents a typical use case where a wide tuning range and a high resolution are desired. Compared to the generated design in 65nm, a higher frequency and a higher resolution are naturally achieved even if the same delay cell is used. To obtain the same minimum frequency, however, more stages are required in the ring oscillator. The monitor can function across a wide VDD range, with $3.13GHz$ max locking frequency achieved at $1.0V$ and $86.3MHz$ at $0.4V$. The last row presents a scenario where the target frequency range is very narrow. In this case, the coarse stage is omitted, and the footprint is minimized. Fig.~\ref{28_apr} shows the automatically generated layouts of the two designs.
\section{Conclusion}
\label{conclusion}

Timing FIAs are powerful attacks that need relatively low efforts to cause devastating consequences. They can be launched through a diverse set of methods and the interference can be highly localized. To mitigate this flexible attack, we propose a distributable and synthesizable timing FIA monitor to detect timing violations caused by clock glitches and data path delay manipulation. The monitor works on the principle of replicating the pulse width of the legitimate clock signal with a configurable delay line to detect injected anomalies. To further reduce development efforts, the FIA monitor is accompanied by an end-to-end automation framework to automatically optimize and implement the monitor based on a few user specifications. 
The chip prototype in 65nm CMOS demonstrates comprehensive and robust detection against all possible types of clock glitches, as well as timing faults induced through voltage, EM, or temperature channels. It self-calibrates to support a wide clock frequency range ($2MHz - 1.26GHz$). 
The small power ($0.487mW$) and area ($0.355MF^2$) enable the monitor to be readily integrated into existing systems in a distributed fashion for optimal coverage of localized attacks. 
Table~\ref{comp} summarizes the performance of the presented FIA monitor and compares it with state-of-the-art designs.


%

\appendices



\ifCLASSOPTIONcaptionsoff
  \newpage
\fi



\def\UrlBreaks{\do\/\do-}
\bibliography{security_lib.bib}
\bibliographystyle{IEEEtran}
%



%

\begin{IEEEbiography}[{\includegraphics[width=1in,height=1.25in,clip,keepaspectratio]{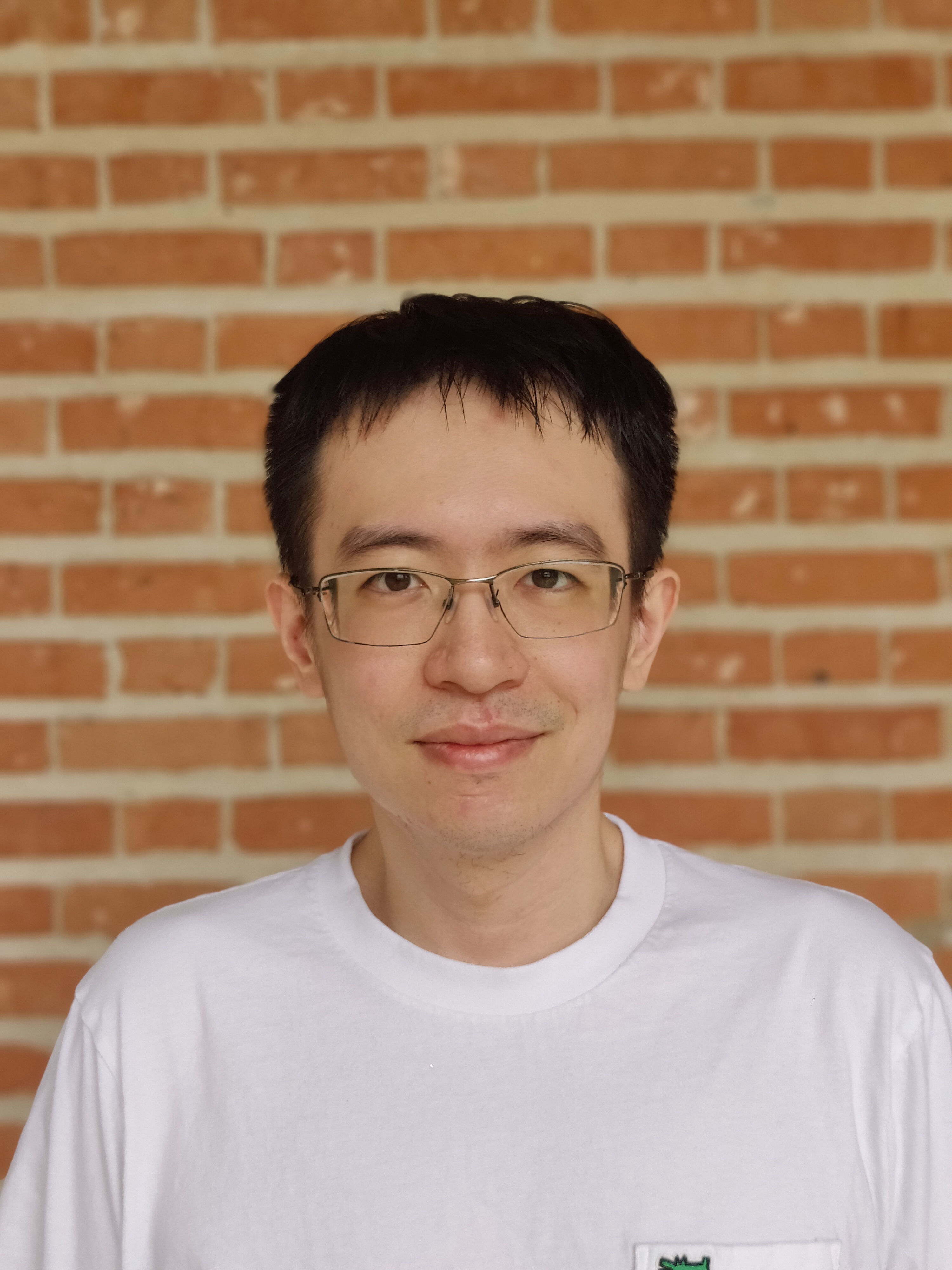}}]{Yan He} received the B.S degree in electronic science and technology from the Zhejiang University, Hangzhou, China, in 2018. He received M.S degree in electrical and computer engineering from Rice University, Houston, USA, in 2022, and Ph.D. degree in electrical and computer engineering from Rice University, Houston, USA, in 2024. His current research interests include analog and mixed-signal integrated circuits design for hardware security.

Currently, he is a research scientist in NVIDIA. He is a recipient of ISSCC Rising Star Award in 2024, SSCS Predoctoral Achievement Award for 2021-2022, and CICC best paper award in 2021.

\end{IEEEbiography}

\begin{IEEEbiography}[{\includegraphics[width=1in,height=1.25in,clip,keepaspectratio]{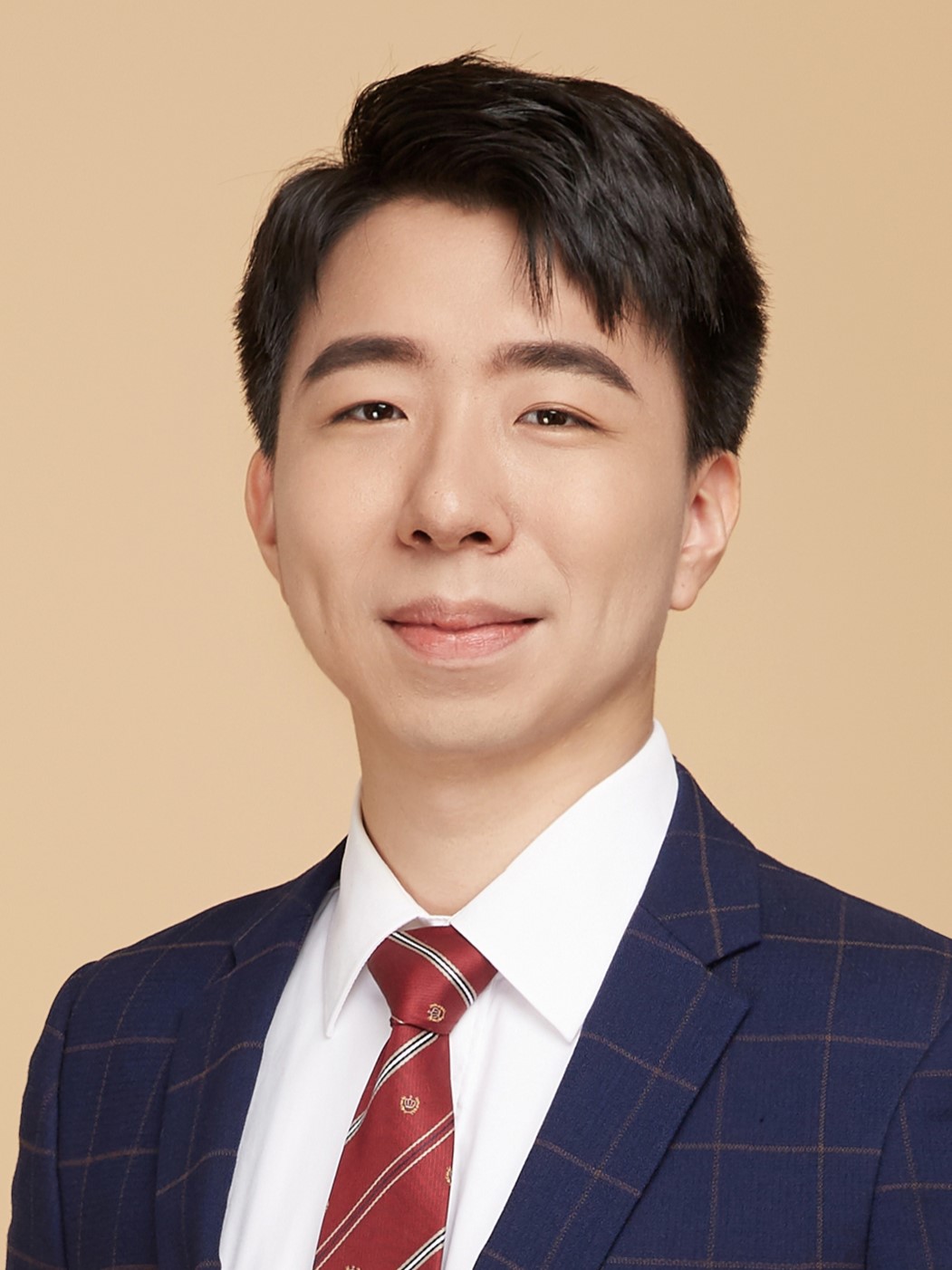}}]{Yumin Su} (Graduate Student Member, IEEE) received his B.S. degree in Electrical and Computer Engineering from Rice University, Houston, TX, USA, in 2023. He graduated summa cum laude with a Distinction in Research and Creative Work. He is currently a 2nd-year Electrical and Computer Engineering Ph.D. student at Rice University, advised by Prof. Kaiyuan Yang. 

His research interests include low-cost hardware security and design automation. 
\end{IEEEbiography}

\begin{IEEEbiography}[{\includegraphics[width=1in,height=1.25in,clip,keepaspectratio]{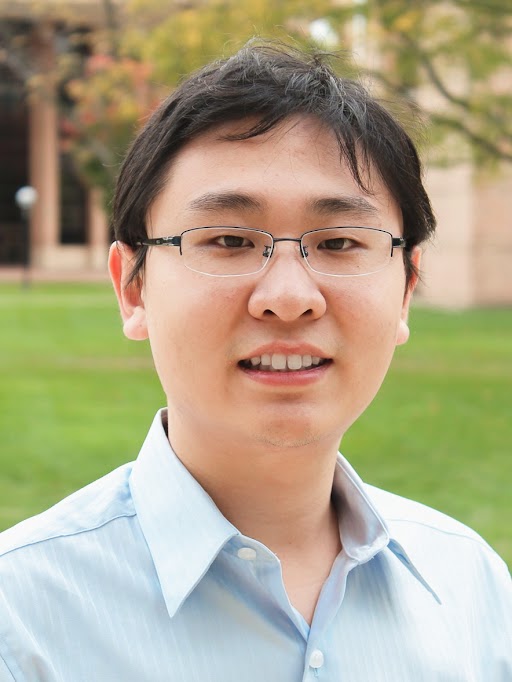}}]{Kaiyuan Yang} (Member, IEEE) received the B.S. 
degree in electronic engineering from Tsinghua University, Beijing, China, in 2012, and the Ph.D. degree in electrical engineering from the University of Michigan, Ann Arbor, MI, USA, in 2017.
 
He is currently an Associate Professor of electrical and computer engineering with Rice University, Houston, TX, USA, where he leads the Secure and Intelligent Micro-Systems (SIMS) Laboratory. His research focuses on low-power integrated circuits and system design for bioelectronics, hardware security, and mixed-signal/in-memory computing.

Dr. Yang is a recipient of the National Science Foundation CAREER Award and the IEEE SSCS Predoctoral Achievement Award. He was also a recipient of the Best Paper Awards from premier conferences in multiple fields, including 2024 Annual International Conference of the IEEE Engineering in Medicine and Biology Society (EMBC), the 2022 ACM Annual International Conference on Mobile Computing and Networking (MobiCom), the 2021 IEEE Custom Integrated Circuit Conference (CICC), the 2016 IEEE International Symposium on Security and Privacy (Oakland), and the 2015 IEEE International Symposium on Circuits and Systems (ISCAS). His research was also selected as the research highlights of Communications of the ACM and ACM GetMobile magazines, and IEEE Top Picks in Hardware and Embedded Security. He is currently serving as an Associate Editor of IEEE TRANSACTIONS ON VLSI SYSTEMS (TVLSI) and a Program Committee Member of ISSCC, CICC, and DAC conferences.
\end{IEEEbiography}



\vfill


\end{document}